\documentclass[5p,times]{elsarticle}
\usepackage{graphicx} 
\usepackage{amsmath}
\usepackage{amsthm} 
\usepackage{amssymb}
\usepackage{stackrel}
\usepackage{mathtools} 
\usepackage{multirow}
\usepackage{url}
\usepackage{epstopdf}
\usepackage{glossaries}
\makeglossaries
\newcommand{\fignm}{Figure}
\newcommand{\fignms}{Figures}
\newcommand{\tablenm}{Table}

\newcommand{\arria}{Arria$^\circledR $ 10 }

\newacronym{alice}{ALICE}{A Large Ion Collider Experiment}

\newacronym{asic}{ASIC}{(Application-Specific Integrated Circuit}

\newacronym{atx_pll}{ATX~PLL}{Advanced Transmit PLL}

\newacronym{bc}{BC}{Bunch Clock}

\newacronym{ber}{BER}{Bit Error Rate}

\newacronym{cern}{CERN}{The European Organization for Nuclear Research (French: Conseil Europ\'{e}en pour la Recherche Nucl\'{e}aire)}

\newacronym{cdc}{CDC}{Clock Domain Crossing}

\newacronym{crorc}{C-RORC}{Common Read-Out Receiver Card}

\newacronym{cru}{CRU}{Common Readout Unit}

\newacronym{cmu_pll}{CMU~PLL}{Clock Multiplier Unit PLL}

\newacronym{ctp}{CTP}{Central Trigger Processor}

\newacronym{daq}{DAQ}{Data Acquisition}

\newacronym{ddl}{DDL}{Detector Data  Link}

\newacronym{diu}{DIU}{Destination Interface Unit}

\newacronym{drorc}{D-RORC}{DAQ Read-Out Receiver Card}

\newacronym{ecc}{ECC}{error correction code}

\newacronym{epn}{EPN}{Event Processing Node}

\newacronym{fec}{FEC}{forward error correction}

\newacronym{fee}{FEE}{Front End Electronics}

\newacronym{fit}{FIT}{First Interaction Trigger}

\newacronym{flp}{FLP}{First Level Processor nodes}

\newacronym{fpga}{FPGA}{Field Programmable Gate Array}

\newacronym{fpll}{fPLL}{Fractional PLL}

\newacronym{fsm}{FSM}{Finite State Machine}

\newacronym{gbt}{GBT}{GigaBit Transceiver optical link}

\newacronym{gol}{GOL}{Gigabit Optical Link}

\newacronym{hb}{HB}{Heartbeat}

\newacronym{hbf}{HBF}{Heartbeat Frames}

\newacronym{hep}{HEP}{High Energy Physics}

\newacronym{hl_lhc}{HL-LHC}{High-Luminosity Large Hadron Collider}

\newacronym{hrorc}{H-RORC}{High Level trigger (HLT) RORC}

\newacronym{ibs}{IBS}{Inter Beam Scattering}

\newacronym{ieee}{IEEE}{Institute of Electrical and Electronics Engineers}

\newacronym{its}{ITS}{Inner Tracking System}

\newacronym{itu_t}{ITU-T}{International Telecommunication Union Standardization sector}

\newacronym{ip}{IP}{Intellectual Property}

\newacronym{jtf}{JTF}{Jitter Transfer Function}

\newacronym{lhc}{LHC}{Large Hadron Collider}

\newacronym{loi}{LOI}{Letter of Intent}

\newacronym{ls2}{LS2}{Long Shutdown}

\newacronym{mft}{MFT}{Muon Forward Tracker}

\newacronym{mgt}{MGT}{Multi-Gigabit Transceiver}

\newacronym{mpo}{MPO}{Multifiber Push-On}

\newacronym{o2}{O2}{Online and Offline}

\newacronym{olt}{OLT}{Optical Line Terminal}

\newacronym{onu}{ONU}{Optical Network Unit}

\newacronym{pci}{PCI}{Peripheral Component Interconnect}

\newacronym{pcie}{PCIe}{Peripheral Component Interconnect Express}

\newacronym{pcix}{PCI-X}{Peripheral Component Interconnect eXtended}

\newacronym{pld}{PLD}{Programmable Logic Device}

\newacronym{pll}{PLL}{Phase Locked Loop}

\newacronym{pfr}{PFR~cycle}{power on/off cycle, firmware upgrade and reset assertion/de-assertion cycle}

\newacronym{pma}{PMA}{Physical Medium Attachment}

\newacronym{pon}{PON}{Passive Optical Network}

\newacronym{prbs}{PRBS}{pseudorandom binary sequence}

\newacronym{prorc}{PRORC}{PCI-based Readout Receiver Card}

\newacronym{pvt}{PVT}{process, voltage and temperature}

\newacronym{rcu}{RCUs}{Readout Controller Units}

\newacronym{rms}{RMS}{Root Mean Square}

\newacronym{rorc}{RORC}{Read-Out Receiver Card}

\newacronym{rf}{RF}{Radio Frequency}

\newacronym{rs}{RS}{Reed-Solomon}

\newcommand{\runthree}{Run 3 }

\newacronym{qcd}{QCD}{Quantum Chromodynamics}

\newacronym{qgp}{QGP}{quark-gluon~plasma}

\newacronym{sda}{SDA}{Serial Data Analyser}

\newacronym{siu}{SIU}{Source Interface Unit}

\newacronym{slink}{S-LINK}{Simple Link Interface}

\newacronym{sfp}{SFP}{small form-factor pluggable}

\newacronym{stf}{STF}{Sub-Time Frame}

\newacronym{trd}{TRD}{Transition Radiation Detector}

\newacronym{tdm}{TDM}{Time Division Multiplexing}

\newacronym{tf}{TF}{Time Frame}

\newacronym{ttc}{TTC}{Timing, Trigger and Control}

\newacronym{ttc_pon}{TTC-PON}{Timing, Trigger and Control (TTC) system based on Passive Optical Networks}

\newacronym{tts}{TTS}{Trigger and Timimg System}

\newacronym{vecc}{VECC}{Variable Energy Cyclotron Centre}

\newacronym{vldb}{VLDB}{Versatile Link Demo Board}

\newacronym{wdm}{WDM}{Wavelength Division Multiplexing}

\pdfoutput=1 

\begin{document}
        
\title{Trigger and Timing Distributions 
using the TTC-PON and GBT Bridge Connection in ALICE for the LHC \runthree Upgrade}

\author[rvt]{Jubin Mitra\corref{cor1}\fnref{fn1}}
\ead{jmitra@cern.ch,jm61288@gmail.com}

\author[rft]{Erno David}

\author[rct]{Eduardo Mendez}

\author[rvt]{Shuaib Ahmad Khan}

\author[rft]{Tivadar Kiss}

\author[rct]{Sophie Baron}

\author[rct]{Alex Kluge}

\author[rvt]{Tapan Nayak}

\address[rvt]{Variable Energy Cyclotron Centre, Homi Bhabha National Institute, 1 / AF Bidhannagar
	Kolkata - 700 064, India}
\fntext[fn1]{Corresponding author.}
\address[rft]{Wigner Research Centre for Physics, Hungarian Academy of Sciences, KFKI, 1121, Budapest, Hungary}
\address[rct]{CERN, CH-1211 Geneva 23, Switzerland}

\begin{abstract}
The ALICE experiment at CERN is preparing for a major upgrade for the third phase of data taking run (\runthree), when the high luminosity phase of the Large Hadron Collider (LHC) starts. The increase in the beam luminosity will result in high interaction rate causing the data acquisition rate to exceed $ 3 $ TB/sec.   In order to acquire data for all the events and to handle the increased data rate, a transition in the readout electronics architecture from the triggered to the trigger-less acquisition mode is required.  In this new architecture, a dedicated electronics block called the \acrfull{cru} is defined to act as a nodal communication point for detector data aggregation and as a distribution point for timing, trigger and control (TTC) information. TTC information in the upgraded triggerless readout architecture uses two asynchronous high-speed serial links connections: the TTC-PON and the GBT. We have carried out a study to evaluate the quality of the embedded timing signals forwarded by the CRU to the connected electronics using the TTC-PON and GBT bridge connection. We have used four performance metrics to characterize the communication bridge: (a) the latency added by the firmware logic, (b) the jitter cleaning effect of the PLL on the timing signal, (c) BER analysis for quantitative measurement of signal quality, and (d) the effect of optical transceivers parameter settings on the signal strength. Reliability study of the bridge connection in maintaining the phase consistency of timing signals  is conducted by performing multiple iterations of \acrfull{pfr}. 
The Intel$^\circledR  $ development kit having \arria \acrshort{fpga} is used for developing the first prototype of the CRU firmware. The test results are presented and discussed concerning the performance of the TTC-PON and GBT bridge communication chain using the CRU prototype and its compliance with the ALICE timing requirements. 
\end{abstract}
\maketitle


\section{Introduction}

\acrfull{alice} at the CERN Large Hadron Collider (LHC) is designed to address the physics of strongly interacting matter and the quark-gluon plasma (QGP) at extreme conditions of temperatures and energy densities. By inclusive studies of 
proton-proton (pp), proton-lead (Pb) and lead-lead (PbPb) collisions at the LHC, ALICE aims to study the properties of QGP.
ALICE has been acquiring data since the year 2009, and has achieved significant milestones and
discoveries since then. An increase in the beam luminosity at the LHC will commence 
from the year of 2020s, which will extend the physics reach of the experiments. 
ALICE will fully exploit the scientific potential offered by this
third phase of LHC data taking run (\runthree) by upgrading the major detector systems, 
associated electronics and data acquisition systems~\cite{abelav2014aliceUpgrade}. 

For \runthree, the collision energy of pp will reach 14~TeV, with maximum instantaneous luminosity of $L = 5\times 10^{34}$cm$^{-2}$s$^{-1}$.
For PbPb collisions, the center of mass energy per nucleon pair, $\sqrt{s_{\rm NN}}$ will be 5.5 TeV, at the 
instantaneous luminosity of $L = 6\times 10^{27}$cm$^{-2}$s$^{-1}$. 
This will correspond to an interaction rate for PbPb collisions of 
50~kHz, compared to the Run2 rate of 8~kHz.
The ALICE upgrade would witness an upsurge in the data volume, with an estimated data flow of
$> 3$~TB/s. Existing trigger based readout architecture is not suitable to cope with hundred times increase in the data taking rate. To handle such data volume, a dedicated data balancing system is introduced in ALICE upgrade of the readout and trigger system \cite{alex2013readout} in the form of a 
\acrfull{cru}. Being at the crossing point of ALICE data streams, CRU manages the aggregation of detector data stream, flow of control requests and distribution of trigger and timing signal information simultaneously.

In this article, we focus on the discussion related to the trigger and timing distribution in ALICE using the \acrshort{cru} framework. 
The integrity of timing signal forms an important technical requirement of the \acrfull{cru} system. 
A detailed study has been performed to ascertain that the multiple high-speed communication links act together by being synchronous to the LHC clock signal to transmit the detector readout data and the timing signals with a constant latency. Moreover, phase information of the embedded clock needs to be preserved and the jitter introduced due to the effect of channel noise also to be retained at a low level. Tests are conducted to confirm the ability of the \acrshort{cru} system to retain the same behaviour with each \acrfull{pfr}.

The trigger distribution system using the \acrshort{cru} is an amalgamation of multi-link technologies involving different protocol standards. For an entire system to operate synchronously and efficiently, the individual designs are optimized to be able to work in coherence to the neighbouring blocks.  In essence, the \acrfull{gbt} and the \acrfull{ttc_pon} link work in conjunction to communicate the \acrshort{ttc} information from the \acrfull{ctp} to a detector through the \acrshort{cru} electronics system. The propagation path involves multiple transition points that involves different protocol conversions, however their concurrent executions might interact subtly. These interactions and their inter-dependencies at juncture points are prone to stochastic fluctuations, and hence proper characterization is needed to affirm the behaviour is deterministic. Hence, piece-wise qualification tests are done, before the final goal to integrate, implement and deploy the design elements. 
The Intel$^\circledR $ \arria development board is chosen as the test board, which carries the same FPGA chip that will be used in the final
CRU development card of PCIe40 \cite{cachemiche2016pcie}.
The advantage of having a development board is that it 
provides easy access to the pins and ports to conduct the signal integrity analysis.

The article is organized as per the following. 
Section~\ref{sec:trigger_to_triggerless} gives an overview of ALICE trigger and the reasoning for trigger-less architecture for \runthree. 
The data flow of the triggerless detector readout raises the need for the use of a new online data frame marker called the heartbeat (HB) trigger, which is also examined in the same section. 
The role of the \acrshort{ttc} systems and the timing distribution to 
different parts of the \acrshort{alice} experiment are outlined in the Section~\ref{sec:clock_distribution}.
High-speed serial links in the TTC communication is discussed in Section~\ref{sec:async_link}.
The flow of detector raw data to the CRU using the GBT link is discussed in Section~\ref{sec:gbt}.
In the Section~\ref{sec:ttc_pon}, the use of a single \acrshort{ctp} link to communicate with the multiple CRUs using a \acrshort{ttc_pon} in a time multiplexing manner is elaborated. 
The symbiosis of the \acrshort{ttc_pon} and the \acrshort{gbt} link technology to form a \acrshort{ttc_pon} and \acrshort{gbt} link bridge made by the \acrshort{cru} firmware is discussed in brief in Section~\ref{sec:ttc_pon_gbt_bridge}.
The design integrates multiple links to detect any unwarranted behaviour in the system and to prevent a failure of cascaded type. Intrinsic system monitoring tools are built to gain statistics about the macroscopic behaviour of the system, which are explained in brief in Section~\ref{sec:design_resilence}.
Section~\ref{sec:results} covers the results related to the latency measurement, the jitter measurement, the \acrfull{ber} measurement and the optimization of transceiver parameters. A discussion of the results is given in Section~\ref{sec:discussion}. Finally, a summary of
the present studies and future outlook 
are presented in the last section.


\section{Triggered and triggerless architecture in ALICE}
\label{sec:trigger_to_triggerless}

The \acrfull{ctp} in ALICE manages the trigger decisions globally and supervise the production of trigger requests by combining the inputs from a system of trigger generating detectors. The CTP plays a pivotal role in identifying rare events, which are recorded for later analysis. 
In the Run2, the \acrshort{alice} uses a hardware trigger strategy, where the signals from minimum bias data sample are selected using thresholds on event multiplicity, 
transverse momenta of tracks and other such observables combining several detectors~\cite{alicePerf2014,alice2014hlt}.
In Run2, the maximum readout rate was limited to 500~Hz for Pb-Pb events. In case the trigger rate exceeds a sub-detector read-out capability, the system saturates and asserts a busy signal. 

In \runthree, the \acrshort{alice} would operate at \textit{six} times the current peak luminosity of $10^{27}~cm^{-2}~s^{-1} $, collecting over a \textit{ten} times the targeted integrated luminosity of $ 1~nb^{-1} $ for the allocated runtime and operating at the collision rate 50~kHz for Pb-Pb ions 
instead of 8~kHz~\cite{abelav2014aliceUpgrade}. 
The physics objective of the upgrade is to improve the precision of the measurement of 
QGP signatures. 
The QGP physics processes do not exhibit signatures that can 
be selected by hardware triggers directly. In the triggerless readout scheme, all events are readout. 
The upgraded event selection strategy uses a combination of triggerless readout scheme and the minimum bias trigger generated from the \acrfull{fit} detector system~\cite{alex2013readout}. 
The new readout architecture for timing and/or trigger distribution topology in ALICE is briefly explained in the Section~\ref{sec:clock_distribution}. 

The principle of the \acrshort{alice} upgrade read-out architecture relies on the \acrfull{tts} ability to efficiently distribute the critical TTC signals with constant latency over optical links to the read-out front-end cards and to receive the busy signal to throttle the trigger distribution when needed. All the data packets originating from the sub-detectors are time tagged. The transmission delay of the read-out data path is not stringent and can use non-constant latency links to the on-line farm. The triggerless data acquisition allows readout of multiple sub-detectors without stressing the trigger decision system when a sub-detector gets busy or faulty. In this manner, the continuous triggerless readout mode increases the event selectivity and allows sampling of the full luminosity. The only drawback that the upgraded system faces is to cope with the massive amount of total generated data that is approximately about 3.6 TB/s. The data flow before the final storage gets reduced by the combined effort of the \acrfull{o2} computing systems. To delimit the overflow of the assembled events across the time frame boundary during data packet formation in the \acrshort{o2} system, a 
new trigger called \textit{heartbeat }~\cite{costa2017readout,Buncic2015online} is defined, as explained in the
next paragraph.

\begin{figure}[!th]
	\centering
	\includegraphics[width=\linewidth]{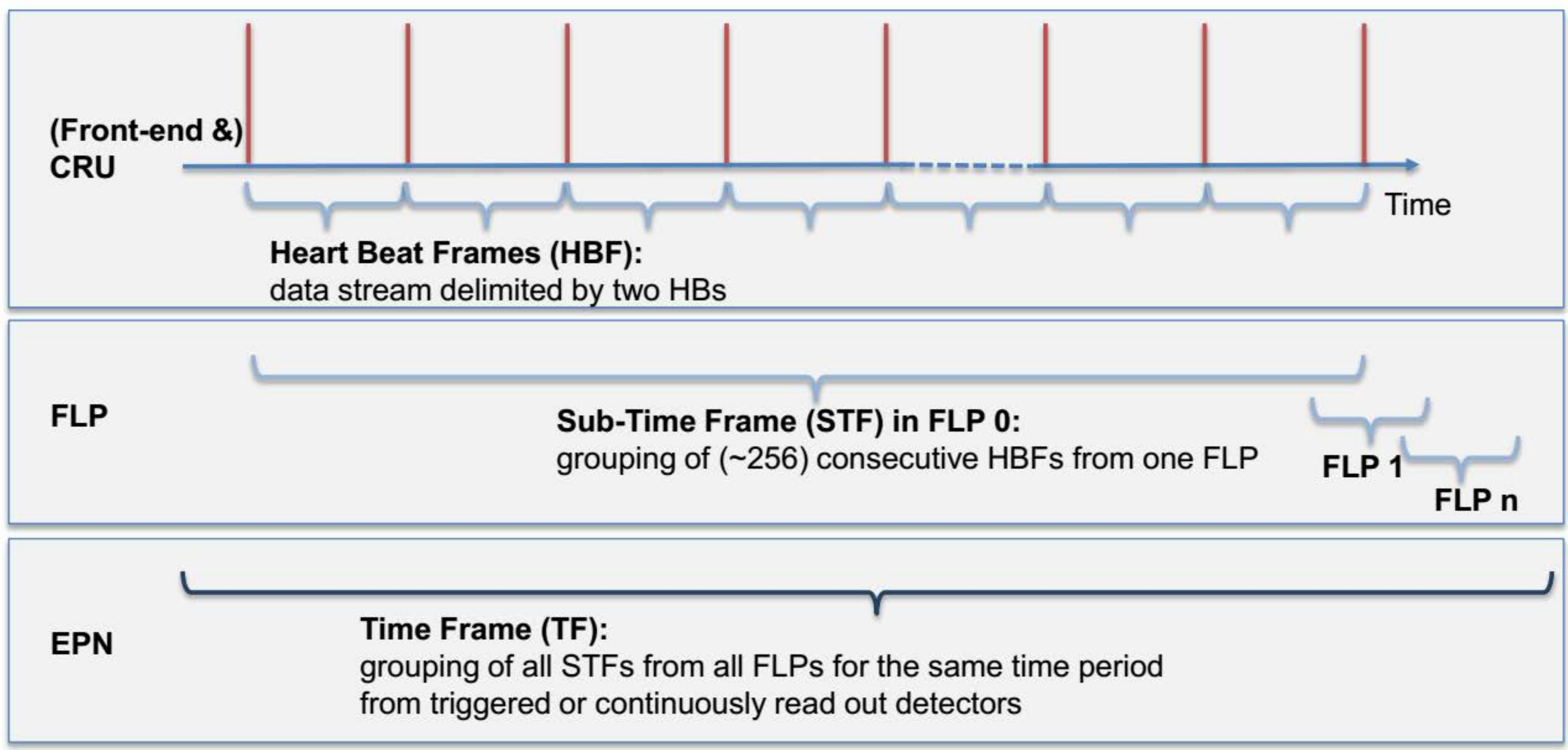}
	\caption{Role of Heart beat trigger in continuous readout mode}
	\label{fig:hb:cont:mode}
\end{figure}

\textit{\acrfull{hb}} in a continuous readout mode
 are asserted periodically to  delimit a stream of readout data~\cite{alice_upgrade_16}. As illustrated in \fignm~\ref{fig:hb:cont:mode}, \acrshort{hb} trigger is used to generate a managable flow of the \acrfull{hbf}.  The \acrshort{hbf}s are forwarded from the readout electronics to the \acrshort{cru} over the \acrshort{tts} links, where it is processed and forwarded to the \acrfull{flp} and then to the \acrfull{epn}, as shown in \fignm~\ref{fig:cru:distribution}. For an each successful \acrshort{hbf} delivery to the \acrshort{flp}, the \acrshort{cru} sends an \acrshort{hb} acknowledge message to the \acrshort{ctp} along with the information about the \acrshort{cru} data buffer.  For the entire operation the \acrshort{lhc} clock is used as a reference signal to synchronize the data flow.
Under a nominal condition, the \acrshort{hb}s rate correspond to one \acrshort{lhc} orbit period of 89.4 $\mu s$ or$\ ~10~kHz$. One \acrshort{flp} accumulates 256 \acrshort{hbf}s to generate a \acrfull{stf} every $\sim 22$~ms, giving a rate of $\sim50~ Hz$. Within the \acrshort{epn} the \acrshort{stf}s coming from all the \acrshort{flp}s are aggregated over the same time period, that includes both triggered or continuously readout detectors to form a complete \acrfull{tf}. Keeping the anticipated customization needed, both the \acrshort{hbf} length and number of the \acrshort{hbf} in a \acrshort{tf} are programmable. 
The \acrshort{hbf}  header, trailer and other IDs are defined to aid the flow managament for the \acrshort{tts} links. One hot encoding is used for the 16 bit trigger codes of the \acrshort{hb} trigger.

\section{ALICE Clock Distribution strategy for \runthree}
\label{sec:clock_distribution}

\begin{figure}[!th]
	\centering
	\includegraphics[width=\linewidth]{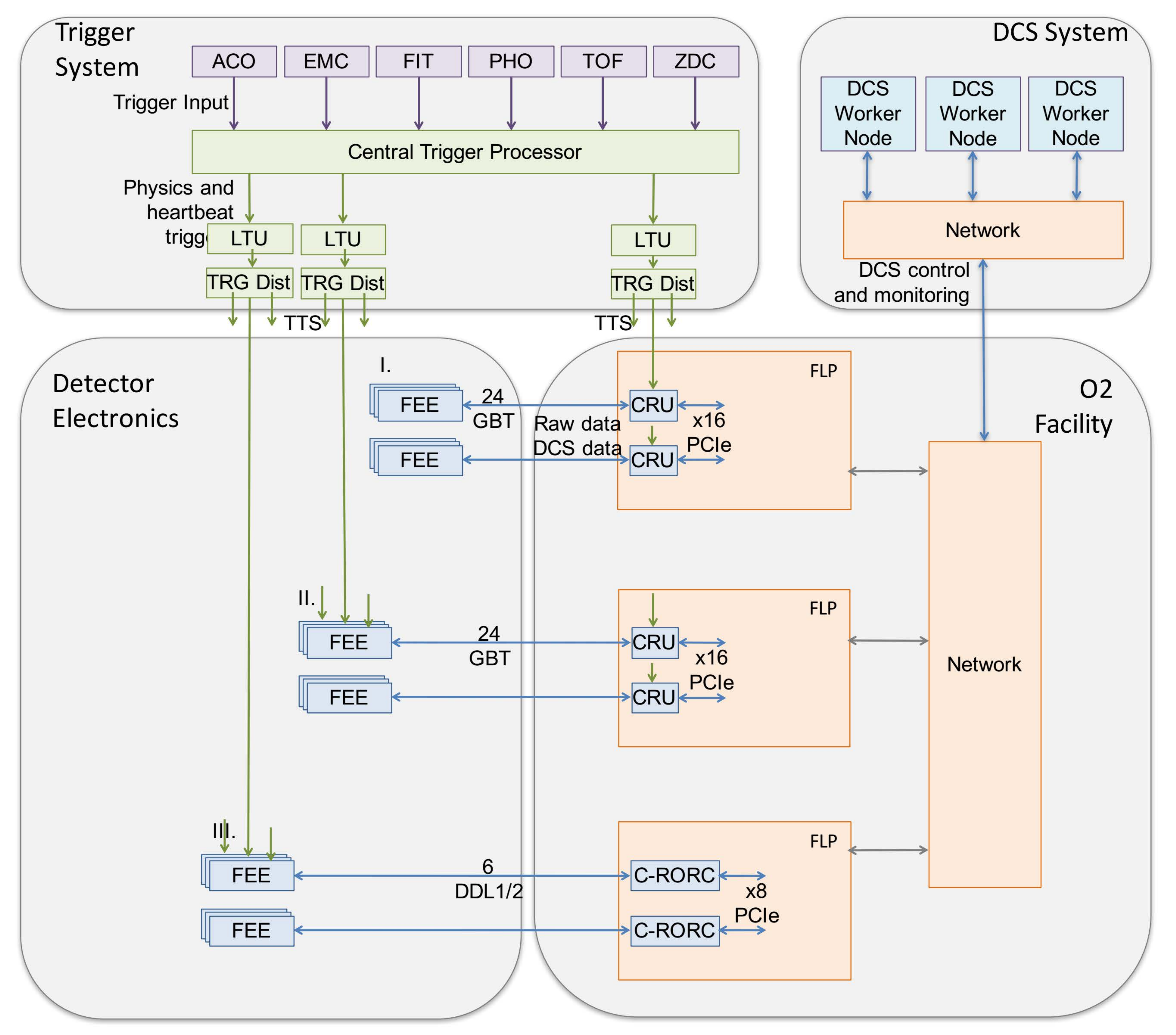}
	\caption{ALICE trigger distribution and detector read-out systems (The TTS distribution path is marked in green).}
	\label{fig:cru:distribution}
\end{figure}

The \acrshort{alice} detector read-out system has three configuration modes to receive the \acrshort{tts} information \cite{alex2013readout} : I. Detectors with non-trigger latency critical systems use the \acrshort{cru} connectivity only, such as the Time Projection Chamber (TPC) and Muon Tracking Chambers (MCH) systems;
 II. Detectors with latency critical trigger information connects directly to the CTP, such as the Inner Tracking System (ITS) ; and 
III. Detectors that would not upgrade to new readout architecture use the \acrshort{crorc} (Run2 readout card) to receive the \acrshort{tts} information on the on-detector electronics via the \acrshort{ttc} protocol. The details of the connectivity of the three
modes are highlighted in the \fignm~\ref{fig:cru:distribution}.

\begin{figure*}
	\centering
	\includegraphics[width=0.7\linewidth]{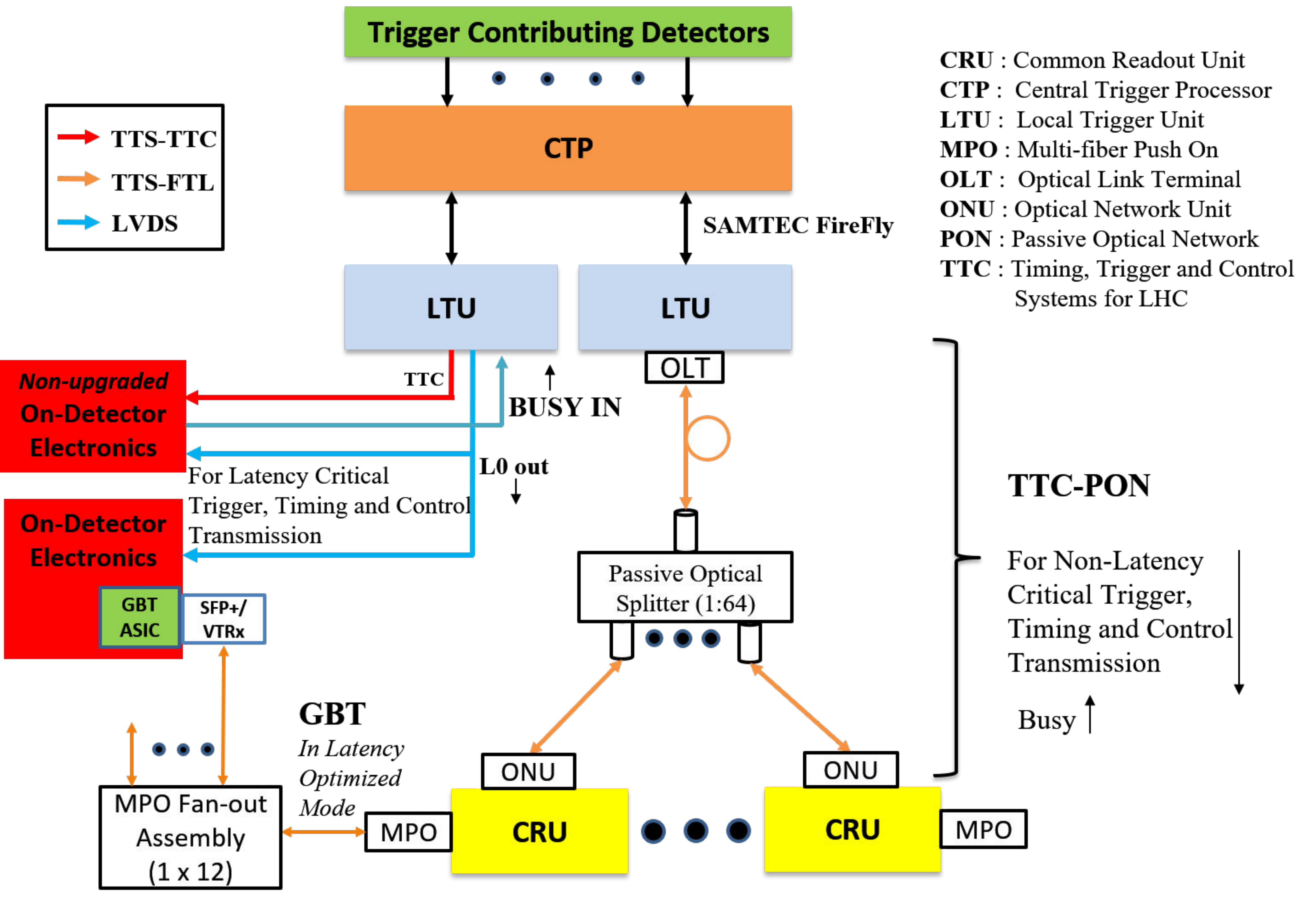} 
	\caption{CRU as Trigger Distribution Unit}
	\label{fig:trigger}
\end{figure*}

The detectors which operate in Type I mode, use the \acrshort{ttc_pon} and \acrshort{gbt} bridge connection to forward the timing information as illustrated in the \fignm~\ref{fig:trigger}.
The detectors can operate in a continuous or a triggered readout mode. Depending on the configuration mode,
the heartbeat trigger or physics trigger is employed.

\section{High-Speed Serial Links in the TTC communication}
\label{sec:async_link}

For a communication system to operate reliably, one of the four classes of the clocking methods are employed, namely the asynchronous (uses no clock signal) scheme, the synchronous (uses same clock frequency and known phase) scheme,  the meso-chronous (uses same clock frequency and unknown phase) scheme and the plesiochronous (uses same clock frequency but with drifting phase) scheme~\cite{Kyung2011jitter}. In the implementation of the asynchronous serial links the clock is embedded within the data stream and behaves in the same manner as the synchronous communication system. The use of embedded clock removes the need for a separate dedicated connection for a clock communication apart from the data stream. However, for the receiver side to recover the embedded clock with efficiency there need to be sufficient transition density in the transmitted bits. Bit transition density is maintained with the help of scrambling algorithm. 

Plenty of commercially viable high-speed asynchronous link standards are available. However, those are not suitable for application in the \acrshort{lhc} environment. The reason being the \acrshort{lhc} operates at a unique frequency of ~40 MHz that is not compatible with the standards of the other commercial links. The use of unconventional clock frequency for data payload communication led the CERN electronics team to develop a custom solution referred as the TTC interface link standards. The TTC standards used in CRU are the \acrshort{gbt} and the \acrshort{ttc_pon}. Comparison between the specification of the two standards are listed in the \tablenm~\ref{table:comparison}. For embedding the clock in the serial link and maintaining the bit transition density, the \acrshort{gbt} uses the block interleaver while the \acrshort{ttc_pon} uses the 8b/10b as channel encoding scheme.

\begin{table}[htbp]
	\centering
	\renewcommand{\arraystretch}{1.2}
	\caption{Showing the basic parametric difference between GBT and TTC-PON}
	\label{table:comparison}
	\resizebox{\linewidth}{!}
	{
		
		\begin{tabular}{|l|c|c|}
			\hline
			\textbf{Parameters} &\textbf{GBT} \cite{gbt09moreira, gbt_baron,amaral2009versatile} & \textbf{TTC-PON} \cite{kolotouros2015ttc}\\\hline
			Technology Specification & Custom & XGPON1 \\
			&&with modifications\\\hline
			Designer Group  & CERN & ITU-T with CERN\\
			&& modifications\\\hline
			Line Rate & 4.8 Gbps & Downstream: 9.6 Gbps\\
			& & Upstream:\phantom{abc} 2.4 Gbps\\\hline
			Payload Rate & 3.2 Gbps & Downstream: $ \sim $7.68 Gbps\\
			& & Upstream:\phantom{abcde} 640 Mbps \\\hline
			Payload Size & 120 bits@40 MHz & Downstream: 192 bits@40 MHz\\
			& & Upstream:\phantom{ab}  $ \sim $16 bits@40 MHz\\\hline
			Wavelength & 850 nm (Multi-mode Tx)& Downstream: 1577 nm\\
			& 1310 nm (Single-mode Tx) & Upstream:\phantom{abc} 1270 nm\\\hline
			Network Topology & Point-to-Point & Point-to-Multipoint\\\hline 
			Encoding & RS Encoding  & 8b/10b\\
			& with Block Interleaver & \\\hline
			Synchronous & Yes & Yes\\
			Trigger Support & &\\\hline
			Trigger Latency & Optical loop-back  & Downstream: $ \sim $100 ns \\
			& Round-trip: ~150 ns &Upstream:\phantom{abc} 1.6 $ \mu $s\\\hline
			Commercially Available & No & No\\\hline
		\end{tabular}
	}
\end{table}

To prevent an anachronistic behaviour in the data packet formation, use of elastic buffers are not preferable. This approach helps to maintain the constant latency in the link that is necessary for arranging the data types having no time-stamp.
Synchronous relationship of flow of events between the source and the receiver over an asynchronous link are preserved by maintaining a certain timing relationship at the physical level of the communication chain. 
An analogy has been given with the distributed systems that behave in a fully synchronous manner only after abiding certain degrees of synchrony \cite{verissimo1997role}. Similarly an asynchronous system can act as a synchronous system provided it abides by certain constraints. In the comparison \tablenm~\ref{table:comparison_synchrony} an attempt has been made to correlate the constraints.
\begin{table}
	\centering
	\renewcommand{\arraystretch}{1.2}	
	\caption{Degrees of Synchrony to behave as a fully synchronous system}
	\label{table:comparison_synchrony}
	\resizebox{\linewidth}{!}
	{
	\begin{tabular}{ll}
		\hline
		\textbf{For Distributed systems} & \textbf{For Asynchronous systems}\\
		(Bounded and known) & (Design Contraints)\\\hline\hline
		
		Processing speed & Frequency locked\\
		Message delivery delay & Fixed latency\\
		Local clock rate drift & Low Jitter\\
		Load pattern & Data Time-stamped\\
		Difference between local clocks &  Phase locked\\\hline
	\end{tabular}
}
\end{table}
The causal relationship of the events are not preserved at the physical level of the protocol stack. Joint use of the clock synchronization and  the syntonization for the recovery of the embedded clock is employed at the physical level. The approach ensures frequency stability with low jitter of the recovered phase locked clock. If the links are operated in latency optimized mode \cite{marin2015gbt} then the latency or the delay path of the data lines remains constant, and is needed for the \acrfull{ttc} data communication. While other levels of complex synchronous information handling like the time-stamp and the trigger management are preserved at the higher level of the \acrshort{cru} firmware logic stack.

\section{GBT Specifications}
\label{sec:gbt}

\begin{figure}[!th]
	\centering
	\includegraphics[width=\linewidth,trim={100 100 50 100},clip]{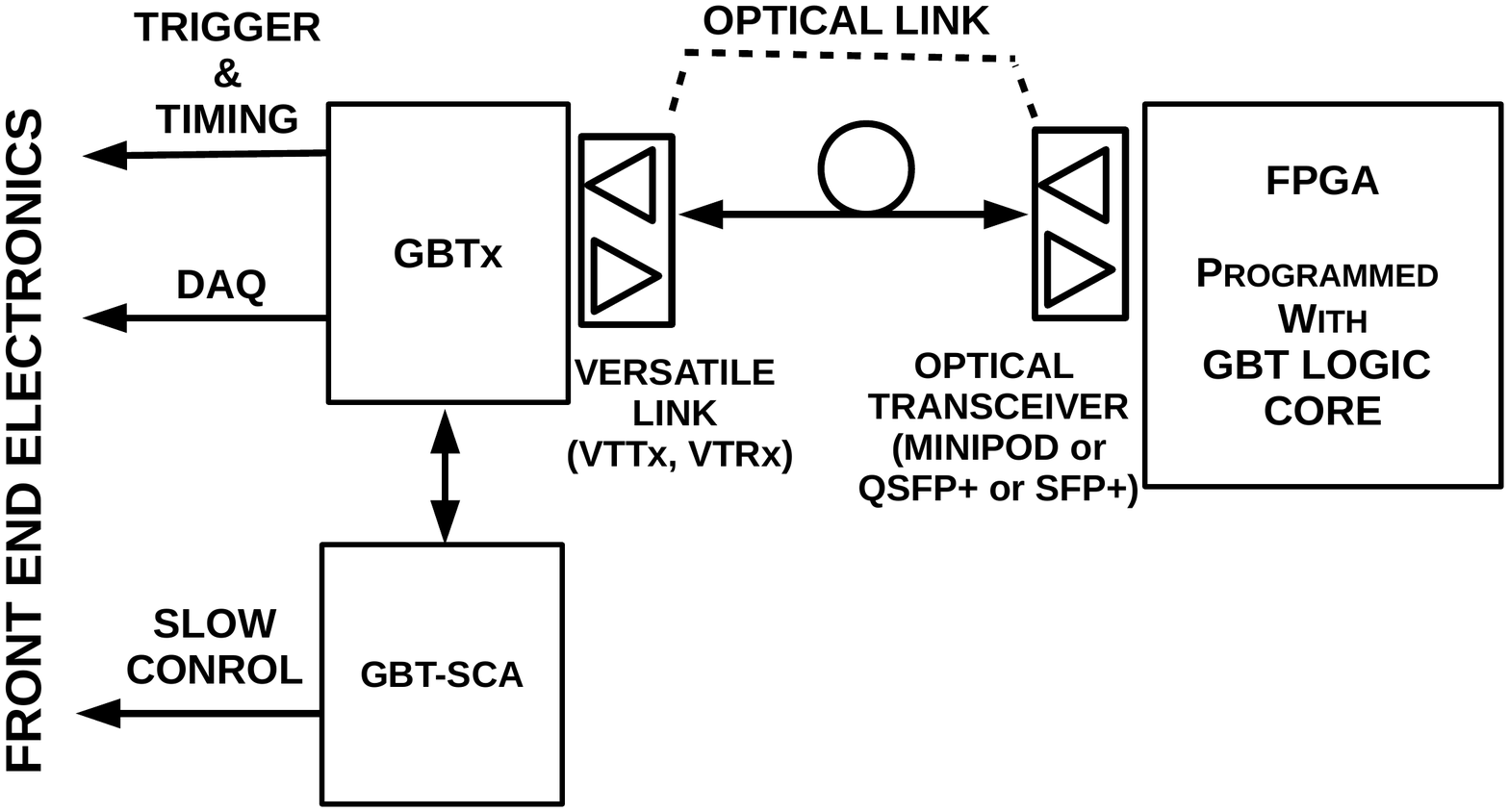}
	\caption{GBT ecosystem}
	\label{fig:gbt_ecosystem}
\end{figure}

The \acrshort{gbt} framework \cite{gbt_baron} defines the technology standards necessary to allow high-speed time critical data communication with high error resilience to communicate reliably from the LHC radiation zone to the readout electronics situated remotely.
  The \acrshort{gbt} ecosystem, shown in the \fignm~\ref{fig:gbt_ecosystem} is composed of three parts, namely, the \acrshort{gbt} \acrshort{asic} that houses the Versatile Link chip along with the \acrshort{gbt} slow control ASIC, Optical Fibre connection operating in a single~mode $ (1310~nm)~$ or a multi-mode $ (850~nm)~$, \acrshort{gbt} the Slow Control ASIC and an \acrshort{fpga} programmed with the \acrshort{gbt} logic core.

\begin{figure}[!th]
	\centering
	\includegraphics[trim={120 120 120 120},clip,width=\linewidth]{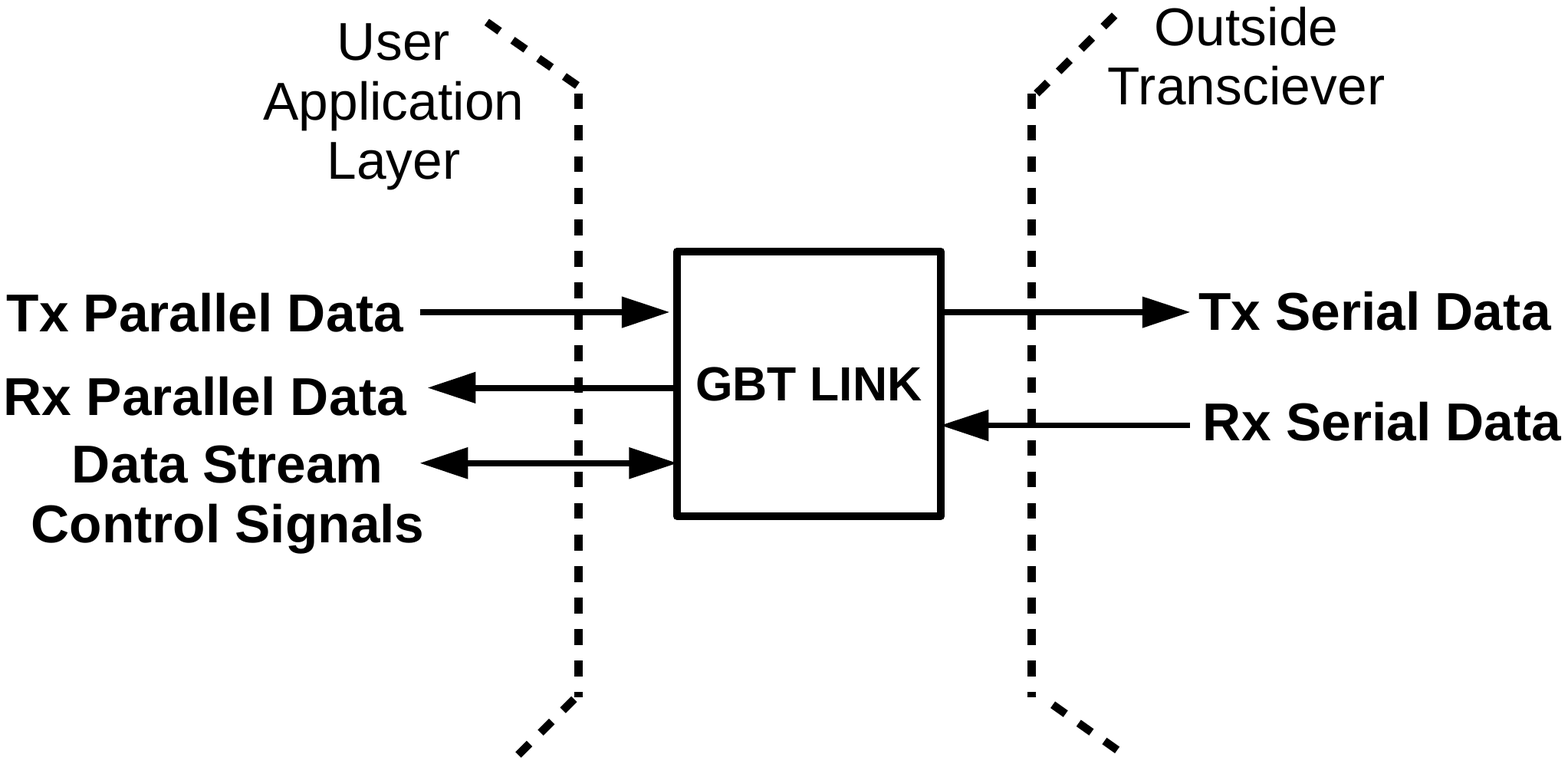}
	\caption{Flow connection of a GBT Link}
	\label{fig:single_gbt}
\end{figure}

The \acrshort{gbt} link supports two modes of operation, the standard mode and the latency optimized mode. The latency optimized mode is needed for a time critical applications that requires constant latency. The \acrshort{gbt} link supports two types of data frame formats, namely the \acrshort{gbt} frame format or the Widebus format. The \acrshort{gbt} frame format appends an error correction code formed from the Reed-Solomon algorithm cascaded with the interleaver and the scrambler. While the Widebus frame format implies that the entire bus is available for the user data and no redundant error checking bits are appended. The line rate of the \acrshort{gbt} link is $4.8~ Gb/s$. For the \acrshort{gbt} scheme the effective data transfer rate is of $ 3.36~Gbps $ and for the Wide-Bus scheme the effective data transfer rate is of $ 4.64~Gbps $. 

In ALICE, maximum 24 \acrshort{gbt} links are required per \acrshort{cru} board. Most part of the \acrshort{cru} \acrshort{fpga} resources are needed for the detector specific logic. To consider a way to save on the GBT specific periphery logic resources a new design approach is required. The effect of optimization on saving the logic resources is studied by Baron et. al.~\cite{baron2009implementing} by sharing one decoder block for several links.
Another level of optimization solution possible in the \arria \acrshort{fpga} that saves on clocking resources and reduces intra-link clock skew is by using a x6 \acrfull{pma} bonded mode~\cite{mitra2016gbt}.  The \acrshort{pma} bonded mode allows \textit{six} \acrshort{gbt} links to be packed closely.
Together the links are referred as the \textit{\acrshort{gbt} Bank}. In other words, the \acrshort{gbt} Bank is that
largest common group formed of the \acrshort{gbt} links that are bonded or clubbed together for an \acrshort{fpga} resource optimization, which is six in this case. The bonded architecture comes with a constraint that all the links need to follow the same standards for a particular \acrshort{gbt} lank. Individually the settings for the \acrshort{gbt} banks are completely configurable. For example, if a designer needs to have 20 links per CRU board then one can split it as three \acrshort{gbt} banks of six links and one \acrshort{gbt} bank with two links.

\section{TTC-PON architecture}
\label{sec:ttc_pon}
Passive Optical Networks (PON) for the particle physics applications at CERN was first proposed in 2009 \cite{pap09pon}. It was later extended for higher speed in 2013 \cite{papakonstantinou2011fully}. The article by Papakonstantinou et. al. (2011) suggested the future use of \acrfull{tdm} in PONs for the \acrshort{ttc} information and the possibility of using \acrfull{wdm} PONs for the \acrfull{daq} applications. Since then the protocol has went through much up-gradation. In our study we have used 2016 version of TTC-PON as shown in \tablenm~\ref{table:comparison}.

\begin{figure}[!th]
	\centering
	\includegraphics[width=\linewidth,trim={30 350 30 80},clip]{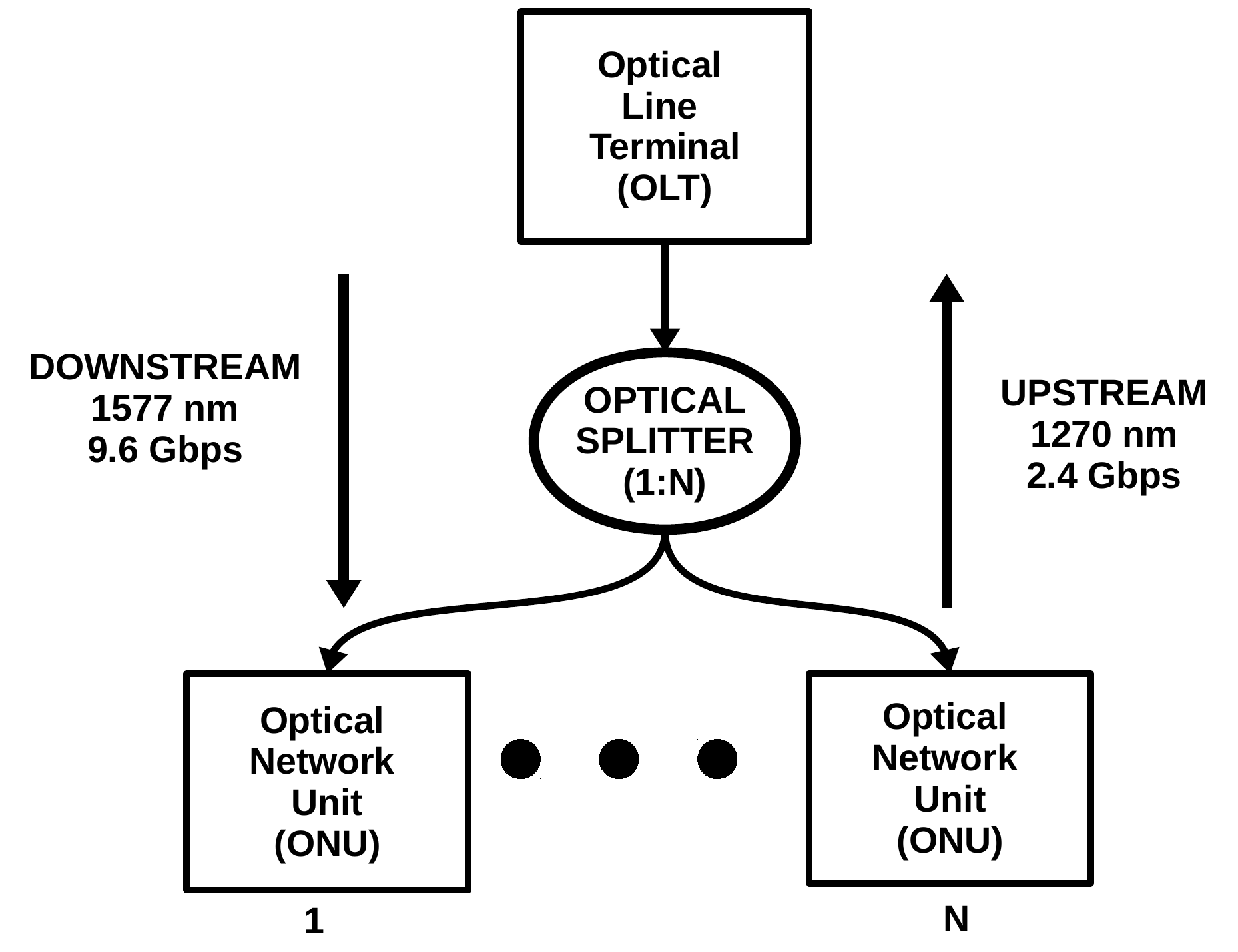}
	\caption{Flow connection of TTC-PON link }
	\label{fig:ttc_pon_arch}
\end{figure}

The TTC-PON architecture is based on PON technology that finds application in Fiber To The Home (FTTH, FTTx) networks. TTC-PON is a single fibre, bi-directional,  point-to-multipoint network
architecture that uses optical splitters to enable a master node or Optical Line Terminal (OLT) to communicate with multiple slave nodes or Optical Network Units (ONUs) \cite{kolotouros2015ttc}, as illustrated in \fignm~\ref{fig:ttc_pon_arch}. 
The downstream (from OLT to ONUs) runs at 9.6 Gbps at operating wavelength band of 1577 nm, while the upstream (from ONU to OLT) runs at 2.4 Gbps operating in wavelength window of 1270 nm.
Using the \acrshort{ttc_pon} technology, the \acrfull{ttc} information from the \acrshort{ctp} are communicated over an optical link in a time multiplexed fashion, that allows a single link to transmit the \acrshort{ttc} information to be splitted among the multiple CRUs  \cite{mitra2016common, mendes2017pon}, as shown in \fignm~\ref{fig:trigger}. The link topology reduces the number of links to be used and hence minimizes the hardware costs involved significantly.  

\section{TTC-PON and GBT bridge for TTC communication}
\label{sec:ttc_pon_gbt_bridge}

\begin{figure}[!th]
	\centering
	\includegraphics[trim={100 550 100 100},clip,width=\linewidth]{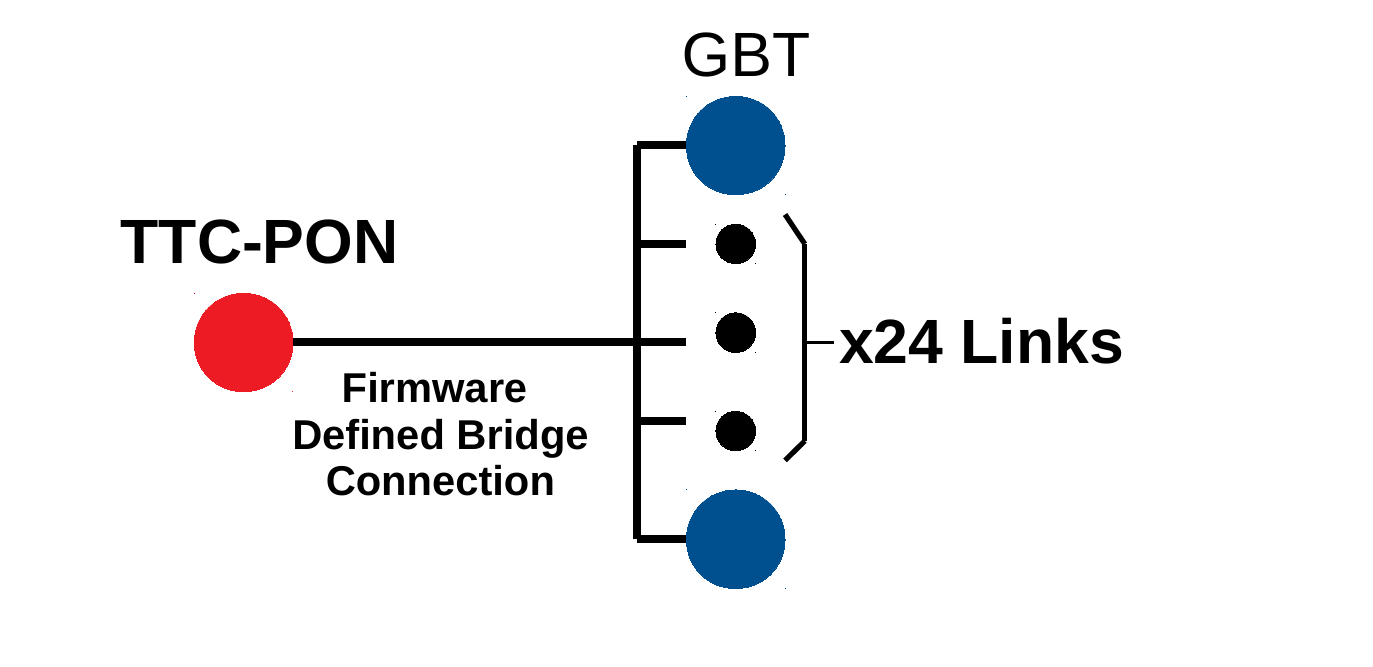}
	\caption{TTC-PON and GBT bridge in star topology connection}
	\label{fig:ttc_pon_gbt_bridge}
\end{figure}

The \acrshort{ttc_pon} and \acrshort{gbt} bridge connection is the interconnection between the two mutually independent \acrshort{gbt} and \acrshort{ttc_pon} links connected using a firmware defined logic. The bridge connection is dedicated for the delivery of the TTC payload. Different types of topology for the bridge connection are possible. For CRU design, the star topology is used for interconnection, where the \acrshort{ttc_pon} forms the central nodal hub for forwarding the TTC information to 24~\acrshort{gbt} links, as shown in the \fignm~\ref{fig:ttc_pon_gbt_bridge}.

\begin{figure}[!th]
	\centering 
	\includegraphics[trim={50 700 50 60},clip,width=\linewidth]{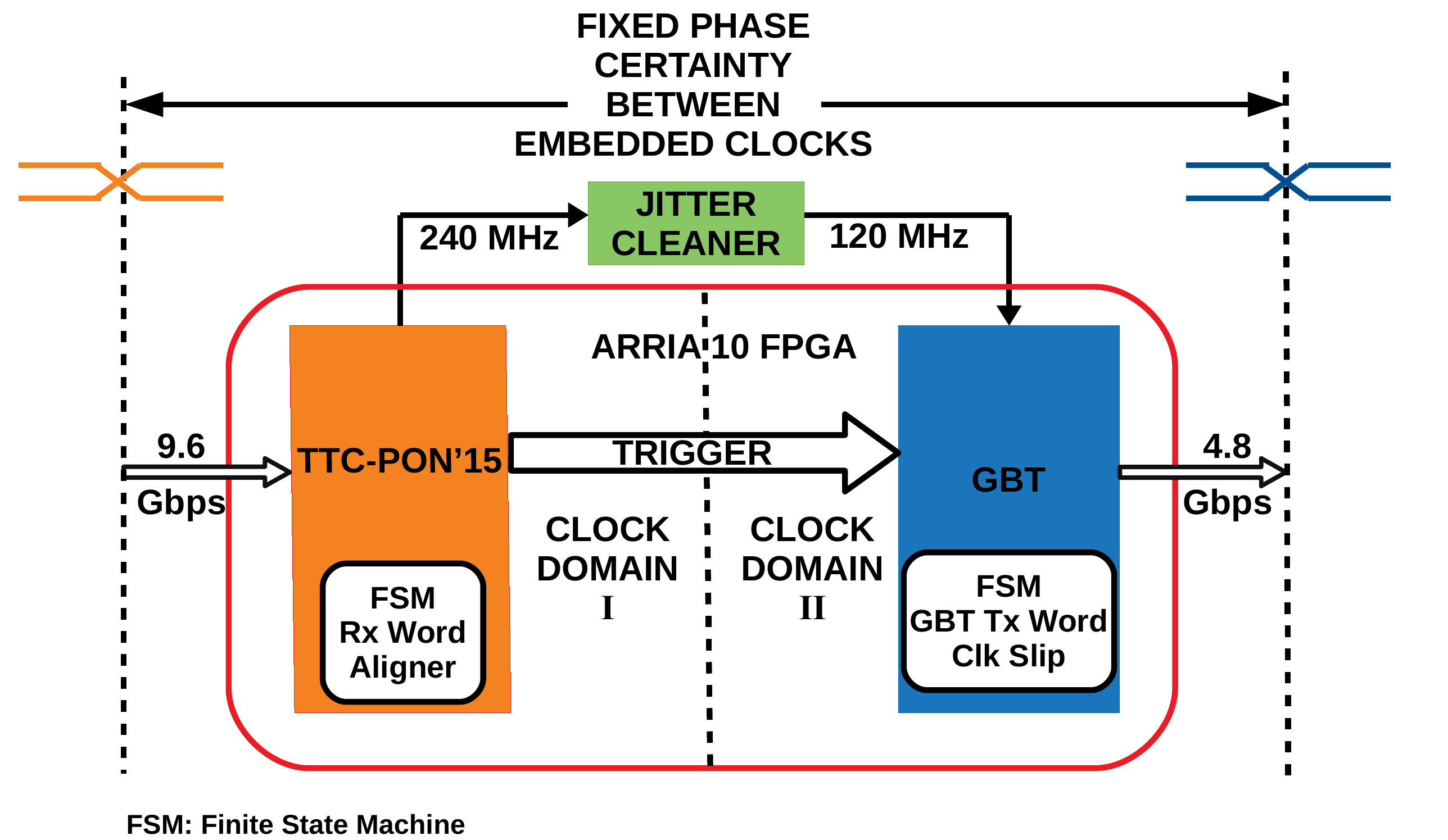}
	\caption{Configuration-I: TTC-PON and GBT bridge connection}
	\label{fig:design:old}
\end{figure}

\begin{figure}[!th]
	\centering 
	\includegraphics[width=\linewidth]{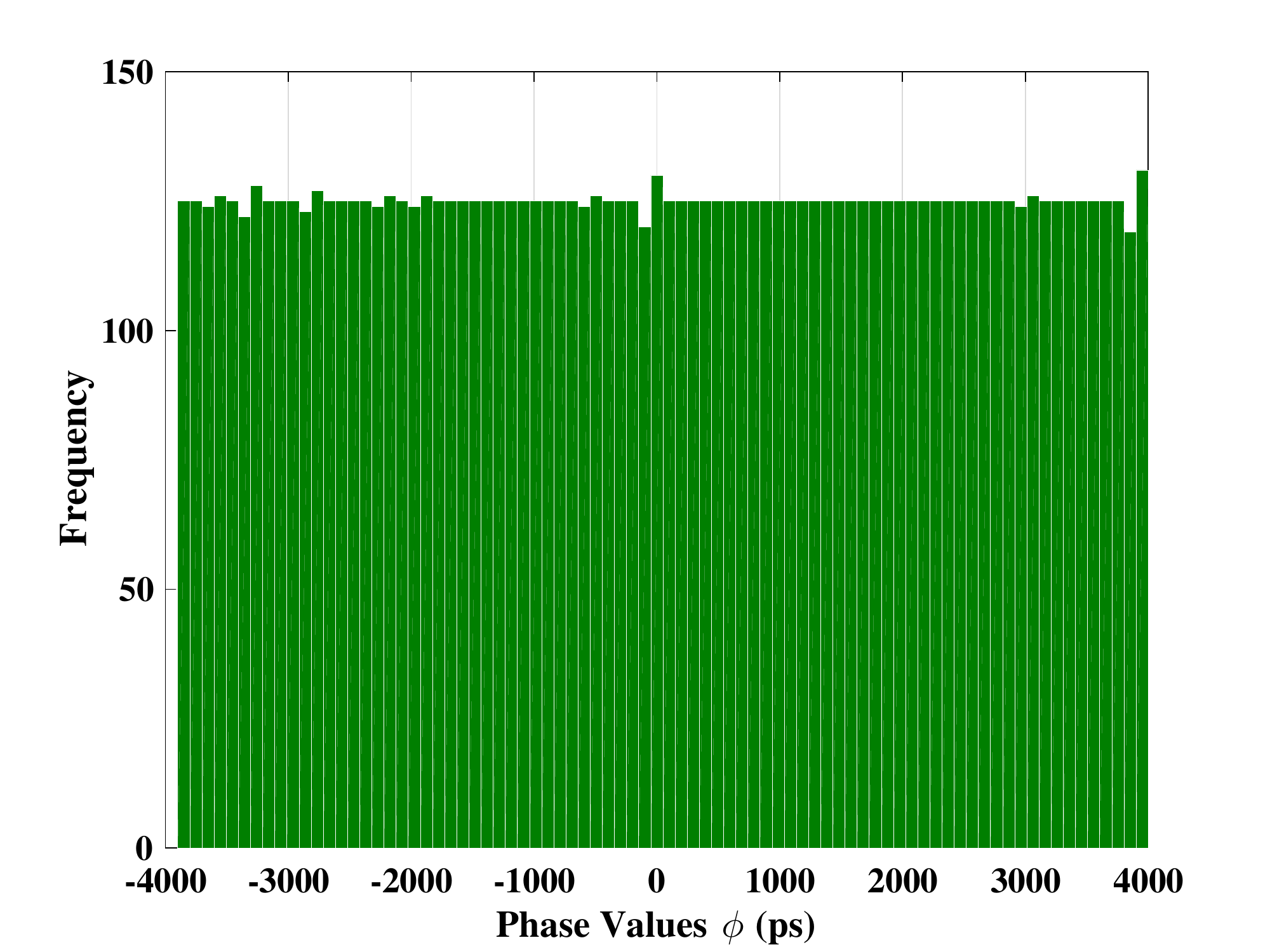}
	\caption{GBT Phase Uncertainty in Tx}
	\label{fig:gbt_phase_uncertain}
\end{figure}

The link connection between a \acrshort{ttc_pon} to the multiple \acrshort{gbt} is elaborated. Initial implementation and testing was based on the scheme where the 240~MHz clock is recovered from the \acrshort{ttc_pon}'15 protocol and then fed into the jitter cleaner before forwarding it to the \acrshort{gbt} at 120~MHz as shown in the \fignm~\ref{fig:design:old} as the configuration-I. However, the implementation on the \acrshort{gbt} side suffered from phase inconsistency of the forward clock with each power cycle or reset cycle. The issue arised because the divider of the \acrfull{mgt} locks the recovered fabric clock at any of the rising edges of the serial clock. In the \fignm~\ref{fig:gbt_phase_uncertain} it is exhibited that for the 10,000 soft reset cycles of the firmware, the phase variation exhibits a uniform distribution over the range of [-4~ns,4~ns]. This has been solved by calibration logic that
 slips the \acrshort{gbt} Tx clock to align with the phase of the recovered clock using 
a \acrfull{fsm}~\cite{mitra2016gbt} .  

\begin{figure}[!th]
	\centering 
	\includegraphics[trim={50 700 50 60},clip,width=0.9\linewidth]{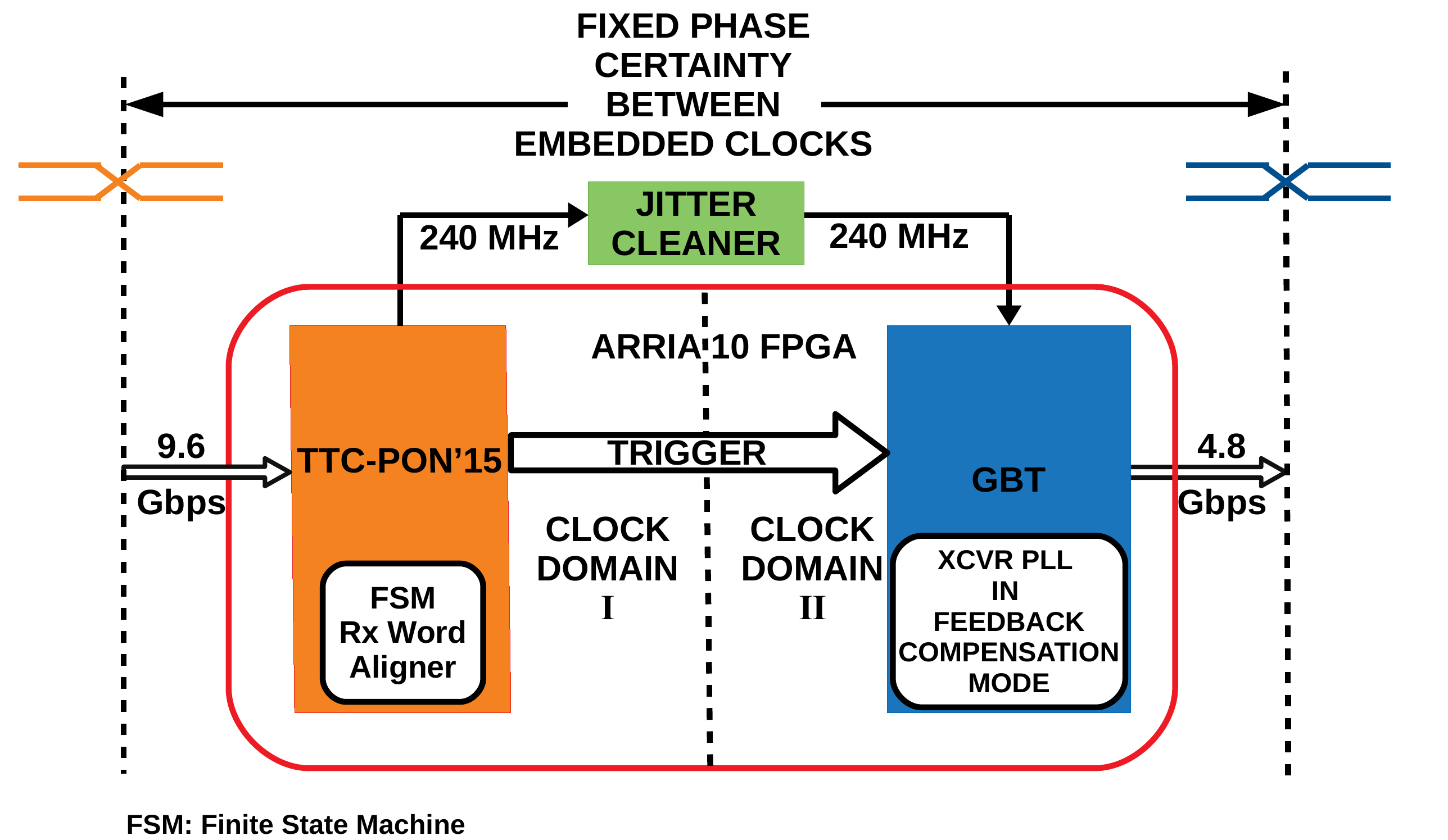}
	\caption{Configuration-II: TTC-PON and GBT bridge connection}
	\label{fig:design:new}
\end{figure}

An improved design option emerged in the version upgrade of the 
Intel~\acrshort{fpga} technology that allows the feedback compensation mode in the transceiver \acrshort{pll} to ensure the deterministic nature of the clock. However, the feature constraints the design solution to operate 240~MHz frequency \cite{native_phy_arria}. Hence, the latest \acrshort{cru} firmware design uses frequency of 240~MHz to cross the entire link chain, instead of stepping down the frequency and stepping up again. The development led to the \acrshort{ttc_pon} and \acrshort{gbt} connection configuration-II as shown in the \fignm~\ref{fig:design:new}. 
In both the configurations given in the  \fignms~\ref{fig:design:old}~and~\ref{fig:design:new} the trigger data path has to cross two clock domains. Even if the clocks are phase locked and of the same frequency, the \acrshort{fpga} sees it coming from two different sources. Hence to avoid the metastability issue in the firmware design synchronizers are added \cite{ste09meta}. Moreover, proper fitter constraints are applied in the firmware logic to lock the logic placements to reduce intra-links skew variation with each firmware upgrade. 


\section{Design Resilience}
\label{sec:design_resilence}

The \acrshort{cru} being a complex heterogeneous system has to deal with multiple links of different communication standards. During a stressful run-time scenario the stochastic fluctuations in the data link pathways might go outside the tolerable zone. The stochastic behaviour is associated with uncertainties and can trigger a chain of cascading upsets in the link chains. Hence, a quantifiable autonomous acquisition system is required to monitor and trace for any unwarranted behaviour. 

As a fall back solution the house keeping tools are included with the main CRU firmware to act as a caretaker to predict any errors and disruption by tracking the macroscopic behaviours of the CRU system. Any deviation of the system behaviour if registered then flags it as a warning or an error to the system management console at the online computing system of the \acrshort{alice}. The inclusive monitoring system involves the three main tools to detect any aberration in the system behavior, as shown in the \tablenm~\ref{table:monitor}. The monitoring system aids in the increase of the resilience and reliability of the system.

\begin{table}[htbp]
	\centering
	\caption{Basic monitoring system to detect macroscopic behaviour}
	\label{table:monitor}
\resizebox{\linewidth}{!}{
\begin{tabular}{rcc}
	\hline
\textbf{Measurement } & \textbf{Monitoring} & \textbf{To}\\
\textbf{Parameter} & \textbf{Signal} & \textbf{Detect}\\\hline
Frequency  & Derived or recovered clock & Presence of Clock\\\\
Phase  & Pico-seconds resolution & PLL and xcvr PHY\\
& phase measurement & working correctly\\
& between the related clocks \cite{mitra2018phase} & \\ \\
Temperature  & The \acrshort{fpga} chip \& the \acrshort{cru} board & Cooling system\\
&& is functional \\\hline
\end{tabular}
}
\end{table}

\section{Results and Discussions}
\label{sec:results}
The entire trigger related logic involves role and functioning of multiple blocks. Each sub-blocks are treated individually, tested, characterized and then integrated in the design system. The test systems are buffeted with various stress scenarios, before compiling the final result in optimum environment condition. 

Since the work deals with timing information transmission, hence tests that give information about the clock quality during a link transition are included. Tests that are entailed in the following sub-sections, are the latency measurement, the jitter measurement, the \acrshort{ber} measurement and the optimization of transceiver parameters.

\subsection{Latency Measurement}
The latency measurement gives an estimation of the logic path delay involved and also senses whether the path traverses through an elastic or an inelastic buffer. Lower the latency is more suitable it is for communication of time-sensitive information, such that service response can be delivered in the shortest period. However, variable latency means the path is ideal for data payload but not for time-critical payloads like in the case of timing and trigger. So, it is the challenge for both the protocols, the \acrshort{ttc_pon} and the \acrshort{gbt}, to meet the low latency with the high throughput and yet be able to preserve the same latency without any variation over the entire run period. Since the serial links traverse through the multiple clock multiplication and the subsequent division zones, the links are subjected to the risk of sudden major variations in the latencies. In the latency measurement also checked for any significant deviations that can perturb the entire time-critical pathway. 

\begin{figure}[!th]
	\centering 
	\includegraphics[width=\linewidth]{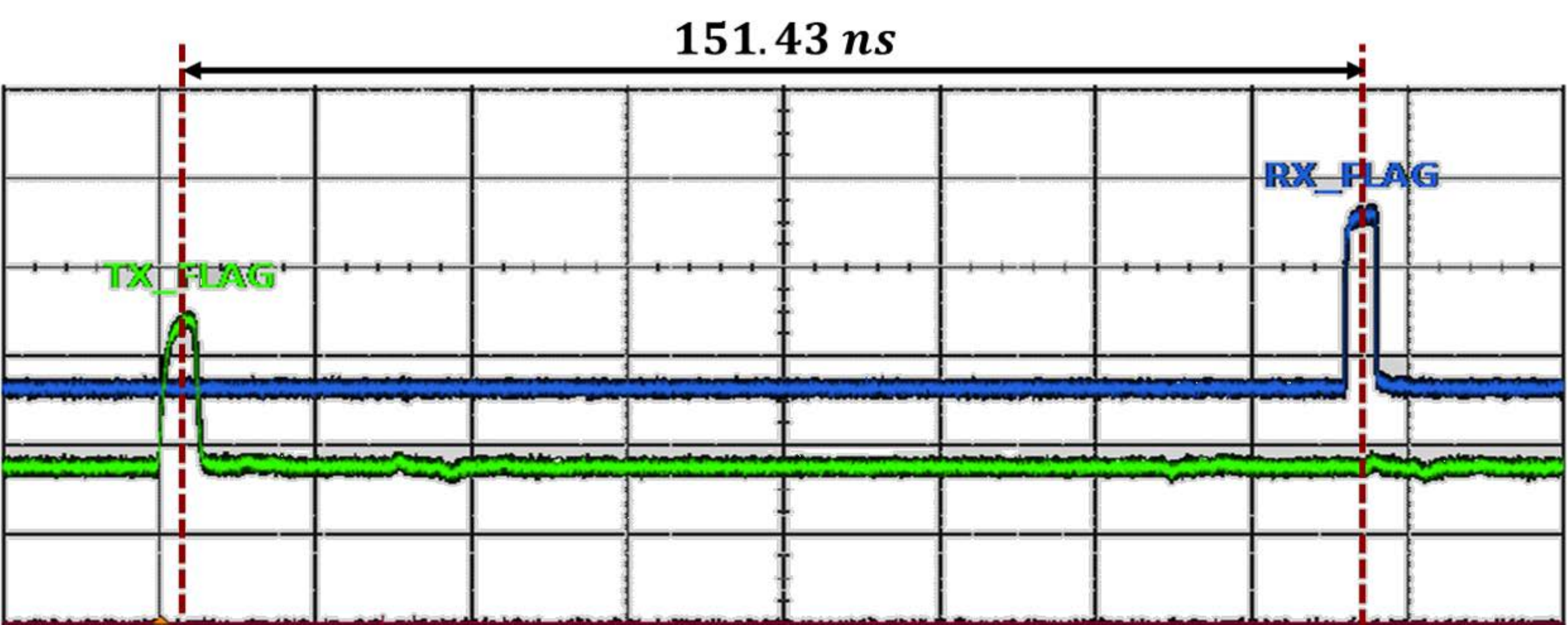}
	\caption{Latency of 151 ns is detected between Transmitting data and Receiving Data}
	\label{fig:latency_flag}
\end{figure}

A special comma is used to measure latency in the \acrshort{ttc_pon}. The comma character is sent every $ 8~\mu s $ (K28.1) and a flag at the \acrfull{olt} level is created to indicate the value is send. A match-flag is created when the \acrfull{onu} received the special character (just after 8b10b decoder). The latency between those two flags are then measured using the oscilloscope.
Several ONU\_RX resets were performed and the position between those flags was deterministic, as shown in the \fignm~\ref{fig:latency_flag}.

\begin{figure*}[!th]
	\centering 
	\includegraphics[trim={100 900 60 100},clip,width=\linewidth]{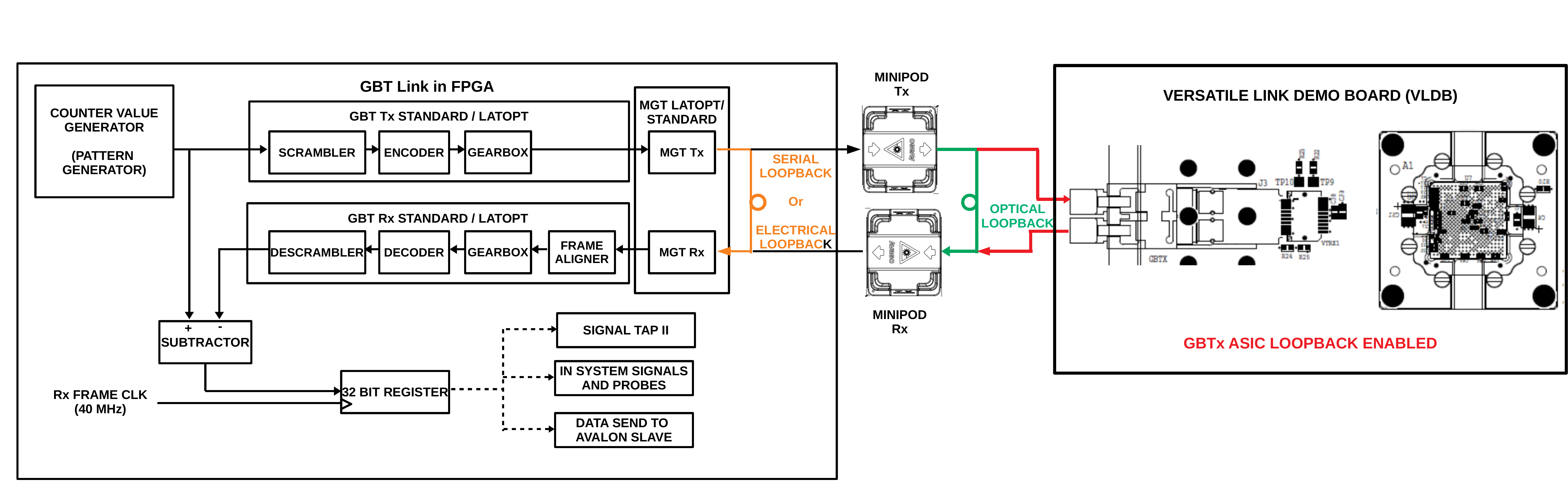}
	\caption{Test Setup for \textit{three} types of round trip delay measurement where each loopback arrangement are marked by different colours}
	\label{fig:lpbk}
\end{figure*}

For the \acrshort{gbt} measurement, an initial version of the PCIe40 DAQ Engine with the \arria \acrshort{fpga} engineering sample is used. Since the setup does not allow to probe at the individual points, an ingenious solution to give a coarse estimate of the latency using firmware based measurement logic is defined. For the measurement a 32-bit ripple counter is used as the generated pattern to communicate over the \acrshort{gbt} stream. The principle of the measurement is to enable the loopback, receive the packet, unwrap from the received \acrshort{gbt} payload and then compare the received counter value with the sender's to estimate the round trip delay. As the \acrshort{gbt} frame arrival rate is of 40 MHz, hence the course measurement of the round trip delay achieved is of $ 25~ns $ resolution.  The round trip delay corresponds to the length of time a signal takes to be sent plus the length of time it takes for the reflected echo of that signal from the receiver to be registered. It includes the serialization and the deserialization time along with the propagation delay. For the measurement as can be seen in the \fignm~\ref{fig:lpbk}, \textit{three} loopback points are chosen, those are : (1) Electrical loopback within the \acrshort{fpga}; (2) Optical loopback; (3) Loopback enabled at the \acrfull{vldb} side.  In the \tablenm~\ref{table:latency} the latency of the \acrshort{gbt} protocol with Tx and Rx configured in the latency optimized mode or the standard mode are tabulated. However, the measurement for the Wide-Bus mode is skipped, as there was no requirement from any ALICE detector group at the time of measurement. The \acrshort{gbt} firmware used for this measurement is the development version used in the year 2015-16 that has got $ 120~MHz $ as word clock.

Both the links exhibited stable latency even when presented with multiple soft or hard resets and power cycles. No unwanted significant deviations are registered.
However, the addition of jitter is present which is a standard behaviour due to the unwanted channel interference and the system noises involved. Details about the jitter measurement of the recovered clock are covered in the following sub-section~\ref{sec:jitter_measurement}.    

\begin{table*}[!th]
	\centering
	\renewcommand{\arraystretch}{1.2}
	\caption{The GBT round trip delay measurements for the multi-level loopback in the PCIe40 }
	\label{table:latency}
	\resizebox{0.85\linewidth}{!}
	{
		\begin{tabular}{lcccc}
			\hline
			\multicolumn{2}{c}{\textbf{GBT Protocol}} & \multicolumn{3}{c}{\textbf{ROUND TRIP DELAY}}\\
			\multicolumn{2}{c}{} & \multicolumn{3}{c}{
				( Resolution of one LHC Bunch clock cycle i.e. 25ns )
			}\\\hline
			\textbf{Transmission Side} &    \textbf{Receiver Side} &    \textbf{Serial or Electrical Loopback}    & \textbf{Optical Loopback} &    \textbf{GBT ASIC Loopback}\\\hline
			\textit{Latency Optimized} &    \textit{Latency Optimized} &    150 ns&    175 ns  &    275 ns\\\hline
			\textit{Latency Optimized} &    \textit{Standard} &   325 ns &    350 ns &    450 ns\\\hline
			\textit{Standard} &    \textit{Latency Optimized} &   200 ns &    225 ns &    325 ns\\\hline
			\textit{Standard} &    \textit{Standard} &    450 ns &    450 ns &    550 ns\\\hline
			
		\end{tabular}
	}
\end{table*}

\subsection{Jitter Measurement}
\label{sec:jitter_measurement}
The asynchronous fast serial trigger links (in the CRU application the \acrshort{gbt} and the \acrshort{ttc_pon}) embed clock signal in the serial data transmission line. The deterministic latency of the timing information transmission is maintained by embedding the synchronization information in the serial data transmission. The embedded clock is recovered by the front end electronics of the detector to generate the \textit{detector specific data packet}. The clock goes through the multiple link transitions, and becomes susceptible to the system and the channel noises. The permissible range of the \acrfull{rms} value of the clock jitter is specific to each sub-detectors requirement and varies with the criticality of the timing of the communication needed. The range of the \acrshort{rms} jitter inclusive of the specifications for all the sub-detectors of the \acrshort{alice}, typically lies within the range of $ 300~ps-20~ps $. 
A preliminary study is conducted to determine if the \acrshort{rms} jitter value of the forwarded clock is less than the lowest jitter requirement of $ 20~ps $.
The test uses the \acrshort{ttc_pon} 2016 version and the \acrshort{gbt} version operating at 240~MHz clock domain. In the Intel$^\circledR $ \arria \acrshort{fpga} devices three types of transmitting \acrshort{pll}s are available: (a) the \acrfull{atx_pll}; (b) the \acrfull{fpll}; (c) the \acrfull{cmu_pll} or the channel \acrshort{pll}. Since the design uses bonded application, hence the \acrshort{atx_pll} and the \acrshort{fpll} can only be used.
The best jitter performance is seen using the \acrshort{atx_pll} over the \acrshort{fpll}. However, the recommendation based on the data rates from the Intel is to use the \acrshort{fpll} for the transmit \acrshort{pll}~\cite{native_phy_arria}, hence in the design fPLL is used.

\begin{figure}[!th]
	\centering
	\includegraphics[trim={15 200 15 25},clip,width=\linewidth]{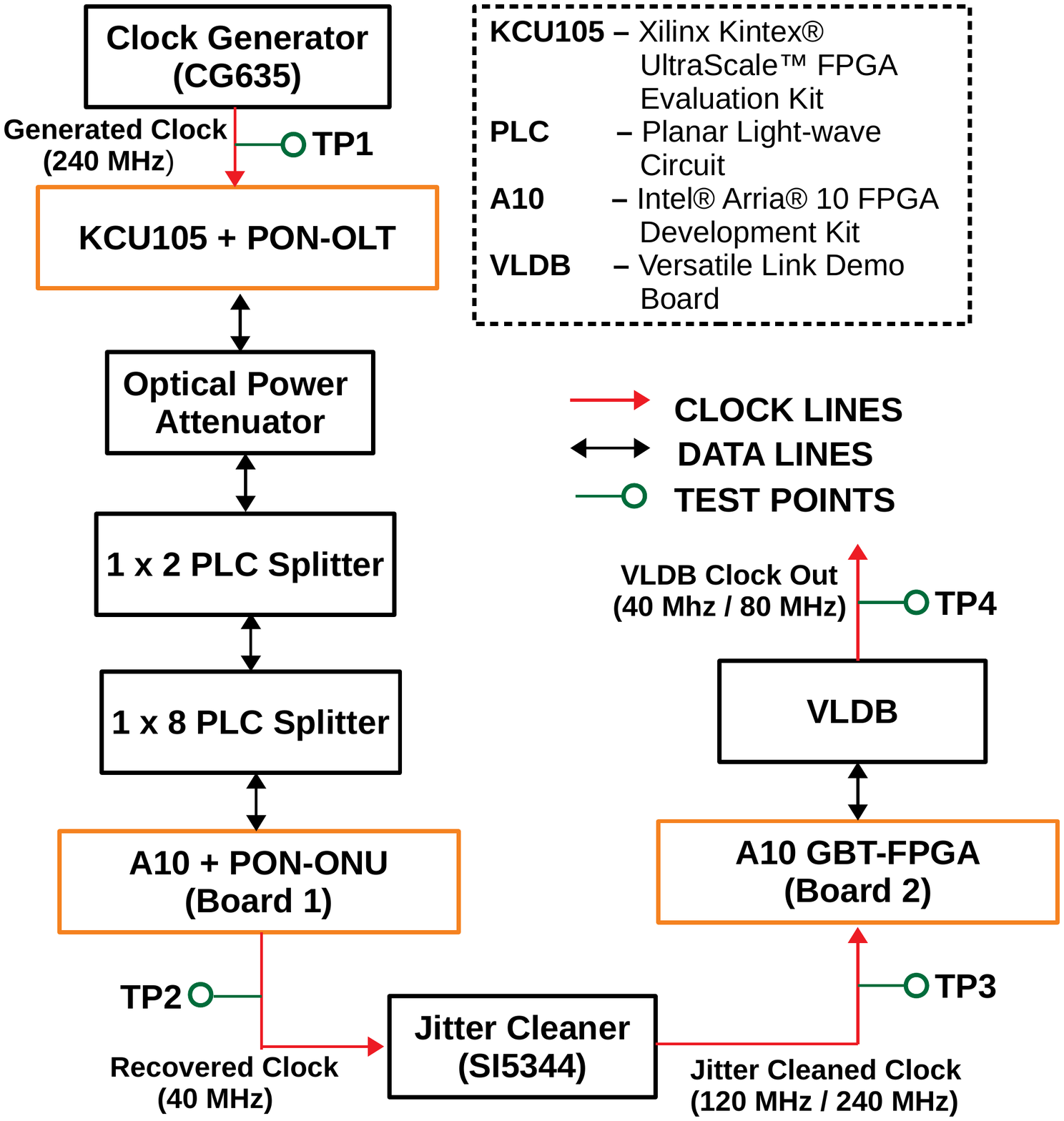}
	\caption{Jitter measurement test setup for the TTC-PON and GBT bridge connection}
	\label{fig:jitter_measurement_setup}
\end{figure}

During the pre-validation test the ideal test scenarios are prototyped with the \acrshort{fpga} development boards having the same family of FPGAs as the final production version. To emulate the \acrshort{ctp} and the \acrshort{cru} hardware, the Kintex  Ultrascale and the \arria development boards are used respectively. For rapid prototyping of the test design, a \textit{split hardware} setup is used to model the behaviour of the \acrshort{cru}. It means that the \acrshort{gbt} protocol is implemented in one \arria \acrshort{fpga} board and the \acrshort{ttc_pon} in other \arria \acrshort{fpga} board, while the clock is transmitted from one board to the other after jitter cleaning it in the SI5344 PLL module. The split hardware model allows an easy access to the \textit{Test Points (TP)} for the jitter measurement test setup as can be seen in the \fignm~\ref{fig:jitter_measurement_setup}. The settings of the configuration parameters of the SI5344 \acrshort{pll} module and the \acrfull{vldb} used for the test is given in the \tablenm~\ref{table:pll:configuration} and the \tablenm~\ref{table:vldb:configuration} respectively. 

\begin{table}[!th]
	\centering
	\caption{SI5344 \textit{Rev B} PLL Configuration}
	\label{table:pll:configuration}
	\begin{tabular}{rl}
		\hline
		\textbf{Parameters} & \textbf{~~~~Status}\\
		\hline \hline
		Free Run Only Mode      &:~~ Not Enabled\\
		Zero Delay Mode         &:~~ Not Enabled\\
		Loop Bandwidth          &:~~ 200 Hz\\
		Fastlock Enable         &:~~ ON\\
		Fastlock Loop Bandwidth &:~~ 1 kHz\\
		Hitless Switching       &:~~ Not Enabled\\
		HoldOff Mode            &:~~ Disabled \footnotemark\\
		Termination &:~~ LVCMOS In-Phase 1.8 V 31 $\Omega$\\\hline
		&  {\scriptsize \phantom{ad} $ ^1 $ REG: 0x052C[0] = 0x0}
	\end{tabular}
	
\end{table}

\begin{table}[!th]
	\centering
	\caption{{VLDB Configuration}}
	\label{table:vldb:configuration}
	\begin{tabular}{rl}
		\hline
		\textbf{Parameters} & \textbf{~~~~Status}\\
		\hline \hline
		Elink Number            &:~~ 0\\
		Channel Number          &:~~ 0\\
		Data Transmission       &:~~ Off\\\hline
	\end{tabular}
\end{table}

\begin{figure}[htbp]
	\centering
	\includegraphics[width=\linewidth]{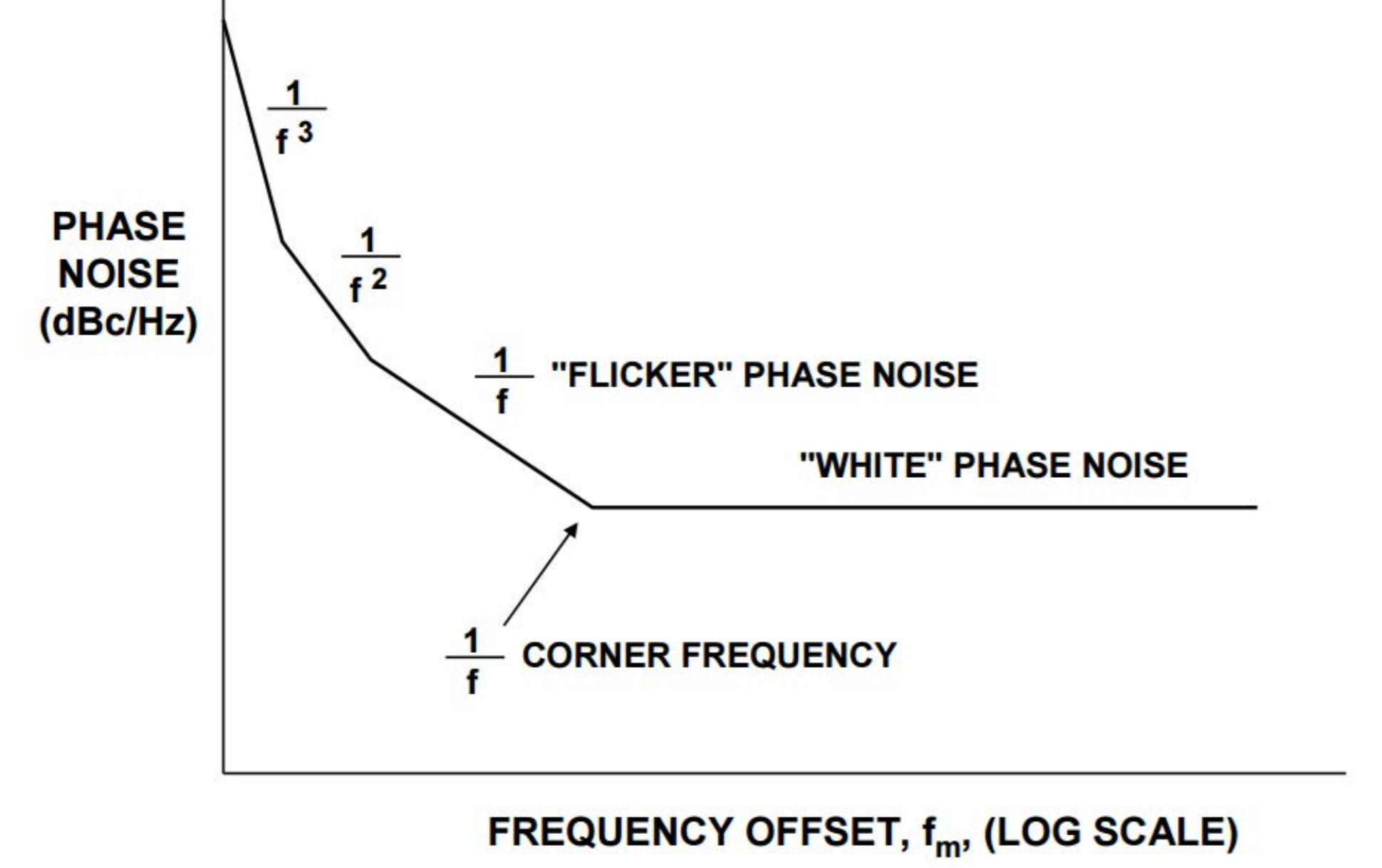}
	\caption{Oscillator phase noise in dBc/Hz vs. frequency offset \cite{Adi2009}}
	\label{fig:ideal_phase_noise}
\end{figure}

The \textit{Phase noise} representation used for the jitter measurement gives an accurate estimation of the phase fluctuations in the frequency domain analysis.
Phase noise is determined as the ratio of the noise in a 1-Hz bandwidth at a specified frequency offset, $f_m$, to the oscillator signal amplitude at frequency $f_O$. The unit used is $dBc/Hz$, where $dBc$ (decibels relative to the carrier) is the power ratio of a signal to a carrier signal, expressed in decibels. 
 It is conventional to characterize an oscillator in terms of its single-sideband phase noise as shown in the \fignm~\ref{fig:ideal_phase_noise}, where the phase noise is in $dBc/Hz$  plotted as a function of frequency offset, $f_m$, with the frequency axis on a log scale. The RMS jitter (in linear terms not dB) is calculated from a piecewise linear integration of the single sideband phase noise data points.
The Equation \ref{eq:rms_jitter:trigger} used for calculation is adapted from the Tutorial MT-008 Analog Devices \cite{Adi2009}. The results are correlated with the Phase Noise Analyzer software generated values. 

\begin{align}
A=Area &= Integrated~Phase~Noise~Power~(dBc),\notag\\
RMS~Jitter &\approx \dfrac{\sqrt{2\times 10^{A/10}}}{2\pi f_O},
\label{eq:rms_jitter:trigger}
\end{align}
where $ f_O $ is the oscillator frequency.

For performing the integration on the \textit{phase noise power values}, the \textit{trapezoidal rule} \cite{Davis2007a} is used over a defined bandwidth  given by the Equation \ref{eq:trapezoidal_rule}.

\begin{align}
\int_a^bf(x)\textit{d}x\approx(b-a)\dfrac{f(b)+f(a)}{2}
\label{eq:trapezoidal_rule}
\end{align}

The experiment calculates the RMS Jitter value from the phase noise power within the bandwidth of \textbf{10 Hz to 20 MHz}. 
As the zone of operation is in the high frequency range, hence the effect of the lower frequency phase noise is neglected \cite{neil2003understanding}. This implies that even though the integrated jitter value computed over the plot A is evaluated higher in relative to the other curve B does not implies that the curve A is better than the curve B. However, if the phase noise curve at the high frequency region of interest for the curve A is below the curve B, then the curve A is considered to be better in performance than the curve B. Such concept is applied in the interpretation of the phase noise measurement plots shown in the \fignms~\ref{fig:phase_noise:pll_choice}, \ref{fig:phase_noise:bw_choice}, \ref{fig:phase_noise:clk_freq_choice}, \ref{fig:phase_noise:jitter_clean} and \ref{fig:phase_noise:link_chain_jitter_addition}. The oscilloscope settings are kept same for the different measurements for the sake of consistency. 

The results contain the measurement of phase noise of clock output from the different test points (TP) in the link chain. The \arria transceiver specific phase noise data points \cite{arria16intel} relative to the reference clock phase noise are also included. 
The \arria transceivers in the Intel data-sheet \cite{arria16intel} gives the phase noise points for the reference clock operating at $ 622~MHz $ frequency.
The REFCLK phase noise requirement at frequencies other than the $622~ MHz$ is calculated using the Equation~\ref{eq:refclkPhaseNoise} adapted from the Intel data-sheet~\cite{arria16intel}.

\begin{align}
&\text{REFCLK phase noise at } f~ (MHz) \nonumber \\
&=  \text{REFCLK phase noise at } 622~ MHz + 20*\log(f/622) \label{eq:refclkPhaseNoise}
\end{align}

The details of the tapping points used during the measurement is given as illustration in the \fignm~\ref{fig:jitter_measurement_setup}. Following measurements are conducted to evaluate the performance and to achieve the best jitter cleaning effect.

\subsubsection{Performance comparison between SI53XX PLL family}
\begin{figure}[!th]
	\centering 
	\includegraphics[trim={60 0 60 20},clip,width=\linewidth]{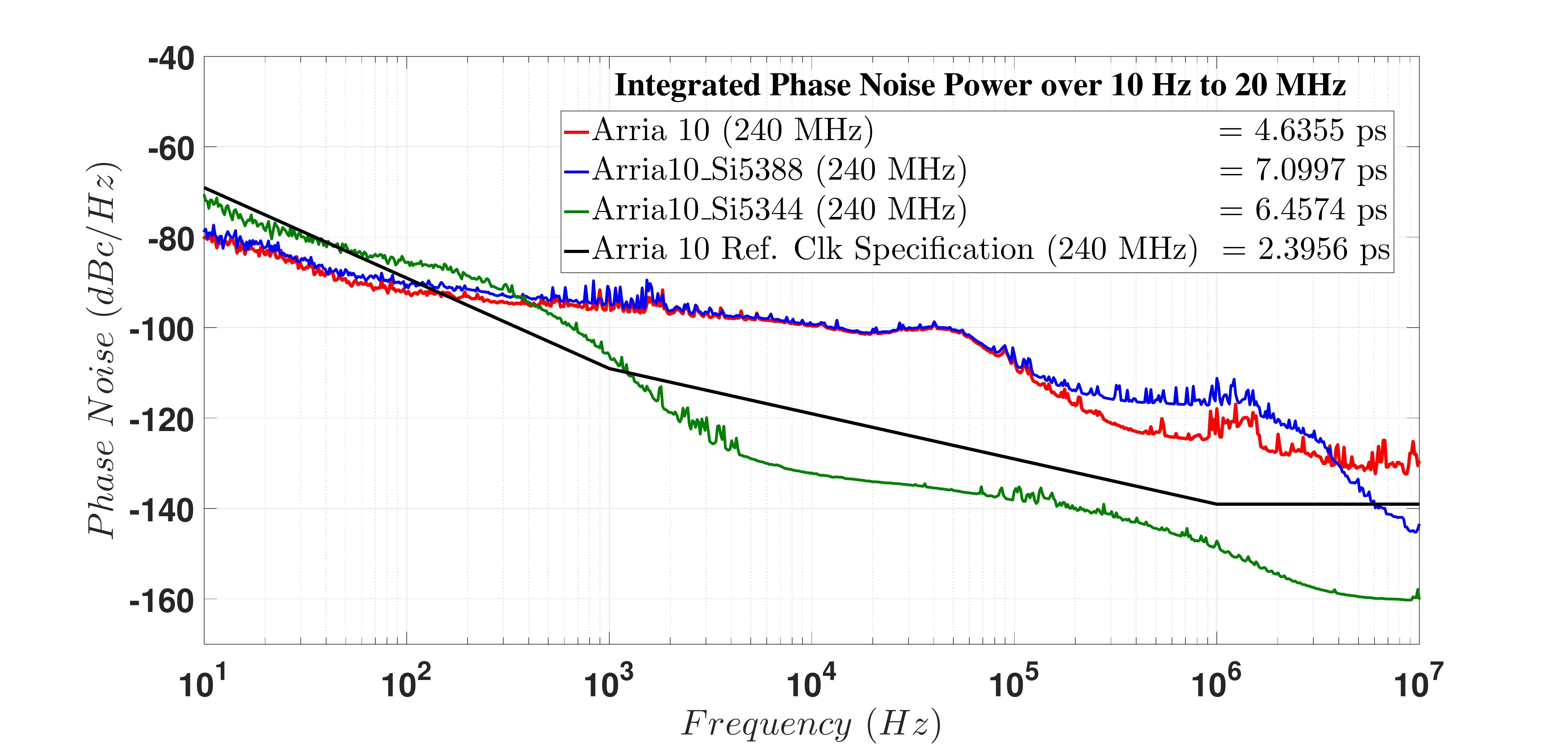}
	\caption{Comparison of jitter cleaning performance between the two jitter cleaners}
	\label{fig:phase_noise:pll_choice}
\end{figure}

 The initial purpose for the study of phase noise measurement is to determine the PLL family that fulfills the CRU requirement. PLLs from the different vendors are characterized by the CERN electronic team members, out of which the PLLs from Silicon Labs SI53XX family found to be suitable for the jitter cleaning requirement for the LHC timing signal. \fignm~\ref{fig:phase_noise:pll_choice} gives comparative result of the jitter cleaning performance of PLLs belonging to the SI53XX family namely the SI5338 and the SI5344. 
The test demonstrates that the SI5344 PLL jitter cleaning is a better match for the phase noise requirement at 240 MHz reference clock frequency for the \arria FPGA SerDes. The test points used for the measurements are TP1 and TP3. The study plays a significant role in deciding the PLL type to be installed in the CRU hardware PCIe40 DAQ Engine.  The SI5345 PLL having 10 output nodes, is a variant of SI5344 PLL family \cite{si5344pll}, which is chosen as the onboard jitter cleaner for the CRU PCIe40 DAQ Engine~\cite{bellato2014pcie}. 

\subsubsection{Performance comparison with PLL bandwidth variation}
\begin{figure}[!th]
	\centering
	\includegraphics[trim={60 0 60 20},clip,width=\linewidth]{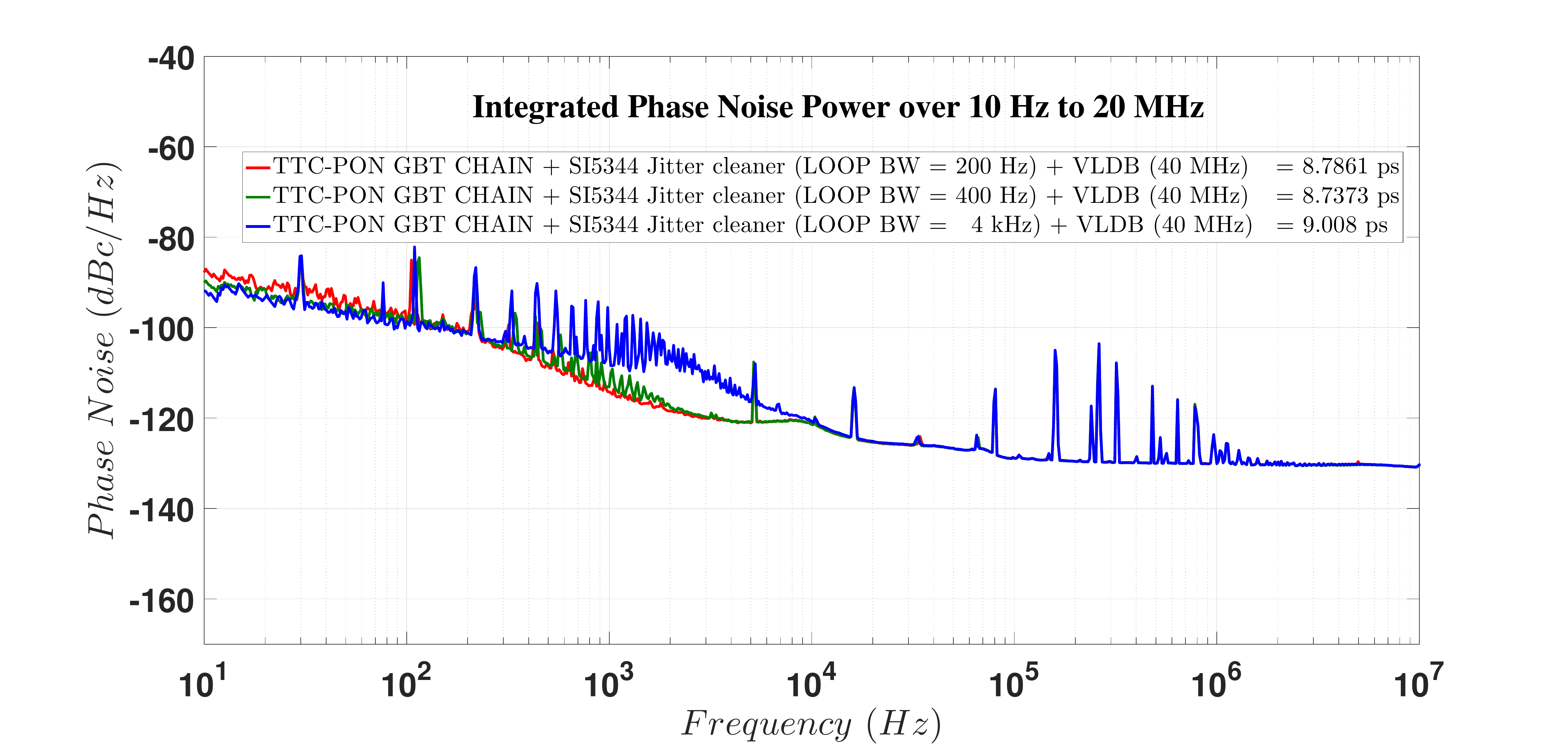}
	\caption{{Phase Noise Analysis (Total Jitter) for VLDB clock output at 40 MHz frequency having the full \nobreak{TTC-PON} and GBT bridge connection with the PLL configured at loop bandwidth 400 Hz and 4 kHz}}
	\label{fig:phase_noise:bw_choice}
\end{figure}

Following test is to determine at which bandwidth configuration the PLL performs at its best. The \fignm~\ref{fig:phase_noise:bw_choice} shows the phase noise study done on the clock signal tapped at the test point TP3. The test is to evaluate the effect of the bandwidth variation on the integrated RMS jitter. From the plot it can be inferred that the PLL gives  best jitter cleaning performance for the 200 Hz bandwidth configuration mode.

\subsubsection{Performance comparison with variation in output clock frequency}
\begin{figure}[!th]
	\centering
	\includegraphics[trim={60 0 60 20},clip,width=\linewidth]{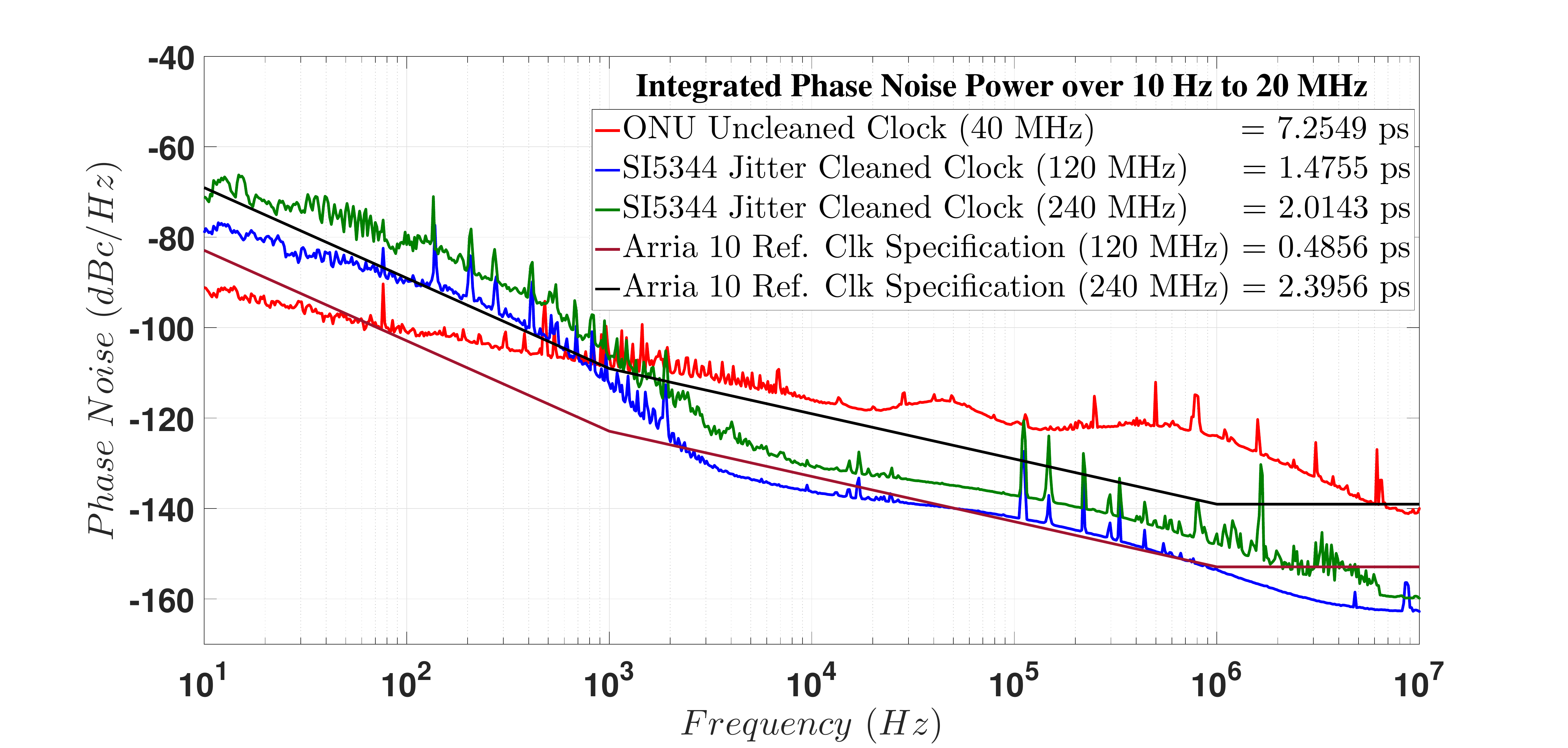}
	\caption{Phase noise analysis (total jitter) for ONU uncleaned clock (40MHz), SI5344 jitter cleaned clock (120 MHz) and SI5344 jitter cleaned Clock (240 MHz)}
	\label{fig:phase_noise:clk_freq_choice}
\end{figure}

After investigating the effect of the operational bandwidth on the jitter cleaning PLL,  the other study that requires attention is the choice of the output clock frequency of the PLL that gives a better jitter attenuation. The CRU firmware design can work at two operational frequencies namely 120~MHz and 240~MHz. The tests with the two frequencies at test point TP3 is shown in the \fignm~\ref{fig:phase_noise:clk_freq_choice}. The \acrshort{cru} firmware with the latest specification uses 240~MHz frequency to make transit of the timing signal from the \acrshort{ttc_pon} to the \acrshort{gbt} without stepping down of frequency at intermediate points. The result shows that the jitter cleaner is able to satisfy the requirement of the \arria \acrshort{fpga} jitter specification of SerDes at 240 MHz reference clock frequency.

\subsubsection{Effect of jitter cleaning performance on integrated GBT and TTC-PON chain}
\begin{figure}[!th]
	\centering
	\includegraphics[trim={60 0 60 20},clip,width=\linewidth]{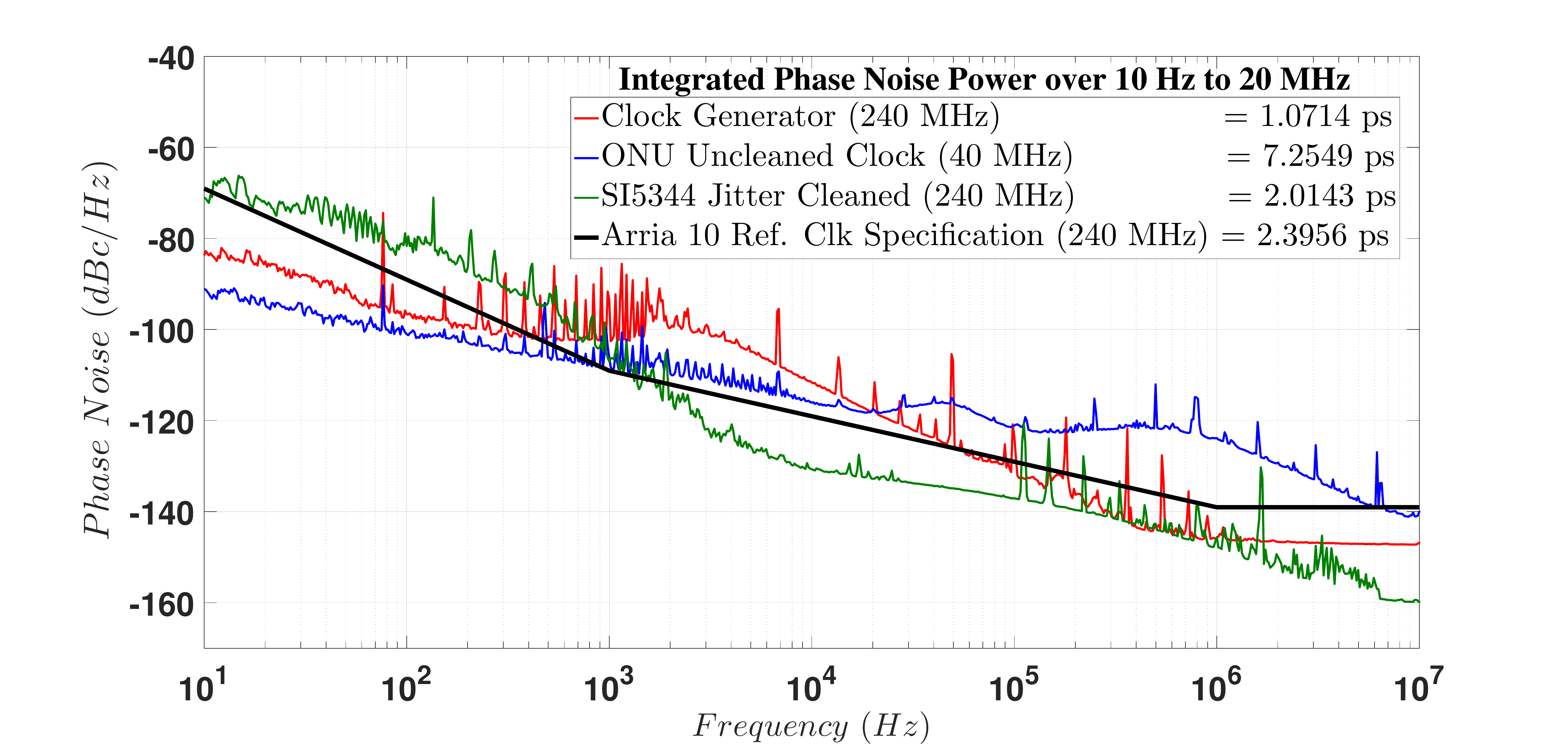}
	\caption{{Phase noise analysis (total jitter) for source clock (240MHz), ONU uncleaned clock (40 MHz) and SI5344 jitter cleaned clock (240 MHz)}}
	\label{fig:phase_noise:jitter_clean}
\end{figure}

The subsequent test for the SI5344 PLL performance  is conducted while fitted in the integrated system. The test points TP1, TP2 and TP3 are used to derive the phase noise plot as shown in the \fignm~\ref{fig:phase_noise:jitter_clean}. The test result validates that the jitter cleaning performance of the PLL keeps the jitter within the tolerable level as specified for the \arria FPGA.

\subsubsection{Comparison of jitter cleaner performance against an ideal test case scenario}
\begin{figure}[!th]
	\centering
	\includegraphics[trim={60 0 60 20},clip,width=0.9\linewidth]{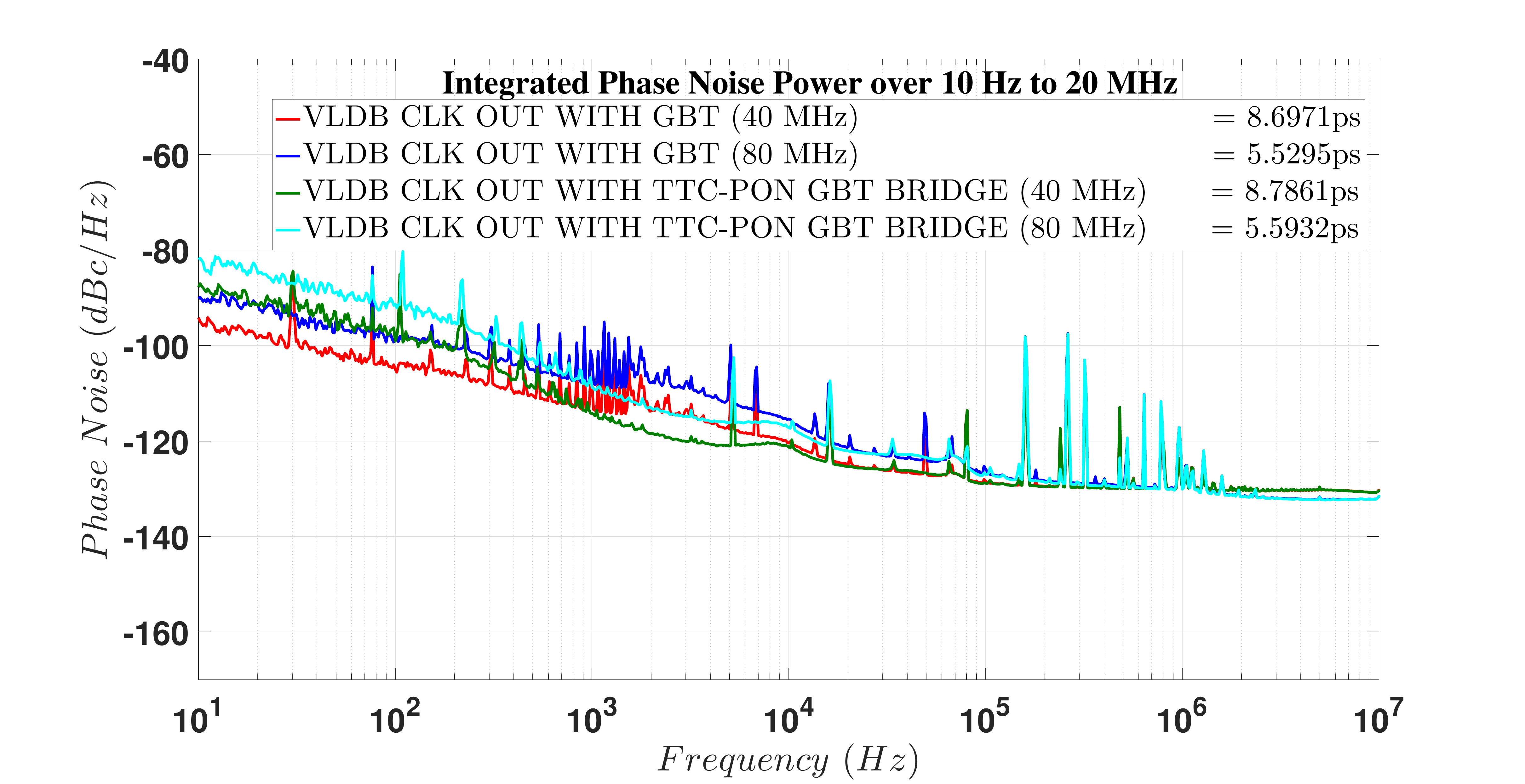}
	\caption{{Phase noise analysis (total jitter) for the VLDB clock output of :  40 MHz o/p with the GBT link connection, 80 MHz o/p with the GBT link connection, 40 MHz o/p with the TTC-PON -- GBT bridge connection and 80 MHz o/p with the TTC-PON -- GBT bridge connection}}
	\label{fig:phase_noise:link_chain_jitter_addition}
\end{figure}

\begin{figure*}[!th]
	\centering
	\includegraphics[width=\linewidth]{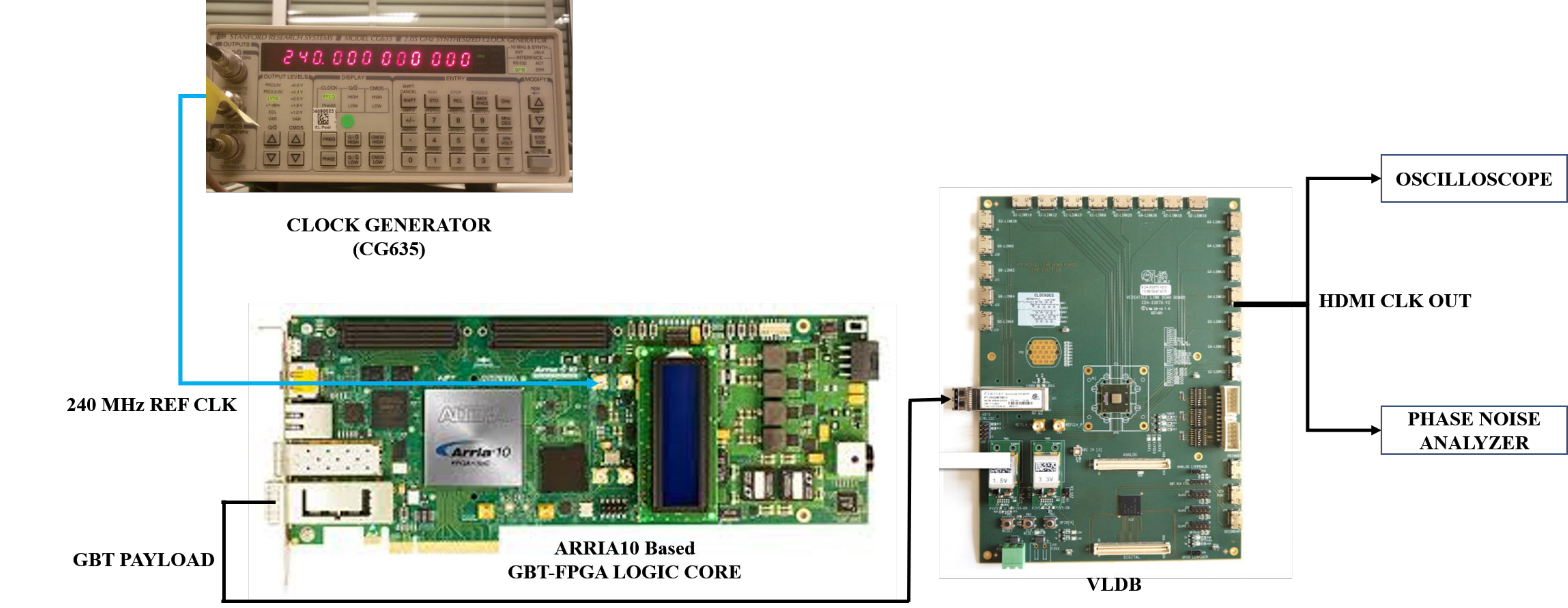}
	\caption{Jitter measurement test setup with GBT CHAIN and VLDB only}
	\label{fig:jitter_measurement_setup_gbt_only}
\end{figure*}

The purpose of the test is to compare the presence of jitter in an ideal experimental condition against the practical scenario with jitter cleaner in use.
The terminal destination of the embedded clock signal in the link chain is the delivery to the GBT chipset. In our test case, we have used VLDB, as it houses the GBT chipset.
The quality of the embedded clock received by the end-point VLDB is studied, where the recovered output clock frequency is set at 40 MHz and 80 MHz respectively. Two types of connection chains are constituted for the test. For the ideal test scenario, all the noisy source points are dropped and the transition points are minimized. It is composed of clock signal originating directly from the clock generator, that gets embedded using the CRU firmware to GBT payload and finally getting communicated to VLDB to be recovered in 40/80 MHz frequency. The setup connection is shown in the \fignm~\ref{fig:jitter_measurement_setup_gbt_only}. For the practical test scenario, the formerly used experimental setup is utilized as shown in the \fignm~\ref{fig:jitter_measurement_setup}.
The results are plotted in the \fignm~\ref{fig:phase_noise:link_chain_jitter_addition} and test result are referred as "VLDB CLK OUT WITH GBT" and "VLDB CLK OUT WITH TTC-PON GBT BRIDGE" respectively. The results of the two test setups show strong positive correlation, hence it can be concluded that the SI5344 jitter cleaner can successfully be employed for the cleaning of the embedded clock during traversal of the TTC-PON and GBT bridge connection.

\begin{table}[!th]
	\centering
	\renewcommand{\arraystretch}{1.4}
	\caption{Comparison of \textit{RMS Jitter} results}
	\label{table:rms_jitter}
	\resizebox{\linewidth}{!}{		
		\begin{tabular}{|l|c|c|c|}
			\hline
			\textbf{CONNECTION} & \textbf{RANDOM}  & \textbf{PERIODIC} & \textbf{TOTAL} \\
			\textbf{TYPE} & \textbf{JITTER} & \textbf{JITTER} & \textbf{JITTER}\\
			& \textbf{(ps)} & \textbf{(ps)} & \textbf{(ps)}\\\hline\hline
			Clock generator (240 MHz) &     0.967 & 1.071 & 1.071\\\hline
			ONU uncleaned clock (40 MHz) &  6.481 & 7.298 & 7.255\\\hline\hline
			
			Si5344 jitter cleaned clock (120 MHz) with loop BW 200 Hz& 1.327 & 1.447 & 1.475\\\hline
			Si5344 jitter cleaned clock (240 MHz) with loop BW 200 Hz & 1.265 & 1.407 & 2.014\\\hline\hline
			
			VLDB clock out with GBT (40 MHz) &        7.409 & 8.697 & 8.697\\\hline
			VLDB clock out with GBT (80 MHz) &        3.501 & 3.884 & 5.529\\\hline
			VLDB clock out with TTC-PON and GBT bridge (40 MHz) &        7.507 & 8.799 & 8.786\\
			having SI5344 loop BW 200 Hz  & & &\\\hline
			VLDB clock out with TTC-PON and GBT bridge (80 MHz) &        3.615 & 4.048 & 5.593\\		
			having SI5344 loop BW 200 Hz  & & &\\\hline\hline	
			VLDB clock out with TTC-PON and GBT bridge (40 MHz)  & 7.439 &       8.738 & 8.737\\
			having SI5344 loop BW 400 Hz & & &\\\hline
			VLDB clock out with TTC-PON and GBT bridge (40 MHz) & 7.727 & 9.018 & 9.008\\
			having SI5344 LOOP BW 4 kHz  & & &\\\hline
	\end{tabular}}
\end{table}

\subsubsection{Comparison of jitter value at intermediate points}
A comparison of all the configurations discussed in the phase noise plots for the integrated RMS jitter is tabulated in the \tablenm~\ref{table:rms_jitter}. The table information is used to derive the \acrfull{jtf} of the individual composite link elements in the link chain.
The \acrshort{jtf} \cite{Schnecker2009Jitter}  is the ratio of the output jitter to the applied jitter on the reference clock, where both the signals are measured as a function of the frequency. The calculated values of the \acrshort{jtf} are given in the \tablenm~\ref{table:jtf}. To summarize the SI5344 PLL satisfies the jitter cleaning requirement of the clock needed in the \acrshort{ttc_pon} and \acrshort{gbt} bridge connection transition.
\begin{table}[!th]
	\centering
	\caption{Comparison of the \textit{RMS Jitter} at each intermediate points in CRU TTC-PON GBT bridge connection}
	\label{table:jtf}
	\resizebox{\linewidth}{!}{
		\renewcommand{\arraystretch}{1.2}
		\begin{tabular}{cccc}
			\hline
			& \textbf{INPUT TOTAL} & \textbf{OUTPUT TOTAL} & \textbf{JITTER TRANSFER} \\
			\textbf{CONNECTION TYPE} & \textbf{RMS JITTER} & \textbf{RMS JITTER} & \textbf{(dB)}\\
			&\textbf{(ps)}   &    \textbf{(ps)}     & \\\hline
			Clock generator & & &\\
			(240 MHz) to ONU & 1.071        & 7.255 & 16.616\\
			Uncleaned Clock (40MHz) & & &\\\hline
			
			ONU Uncleaned Clock & & &\\
			(40MHz) to Jitter Cleaned & 7.255 & 2.014 & -11.132\\
			Clock (240 MHz) & & &\\\hline
			
			Jitter Cleaned Clock  & & &\\
			(240 MHz) to VLDB Clock & 2.014 & 8.697 & 12.706\\
			Out (40 MHz)     & & &\\\hline
		\end{tabular}
	}
\end{table}


\subsection{BER Measurement}
\begin{figure}[!th]
	\centering
	\includegraphics[trim={0 300 0 150},clip,width=0.95\linewidth]{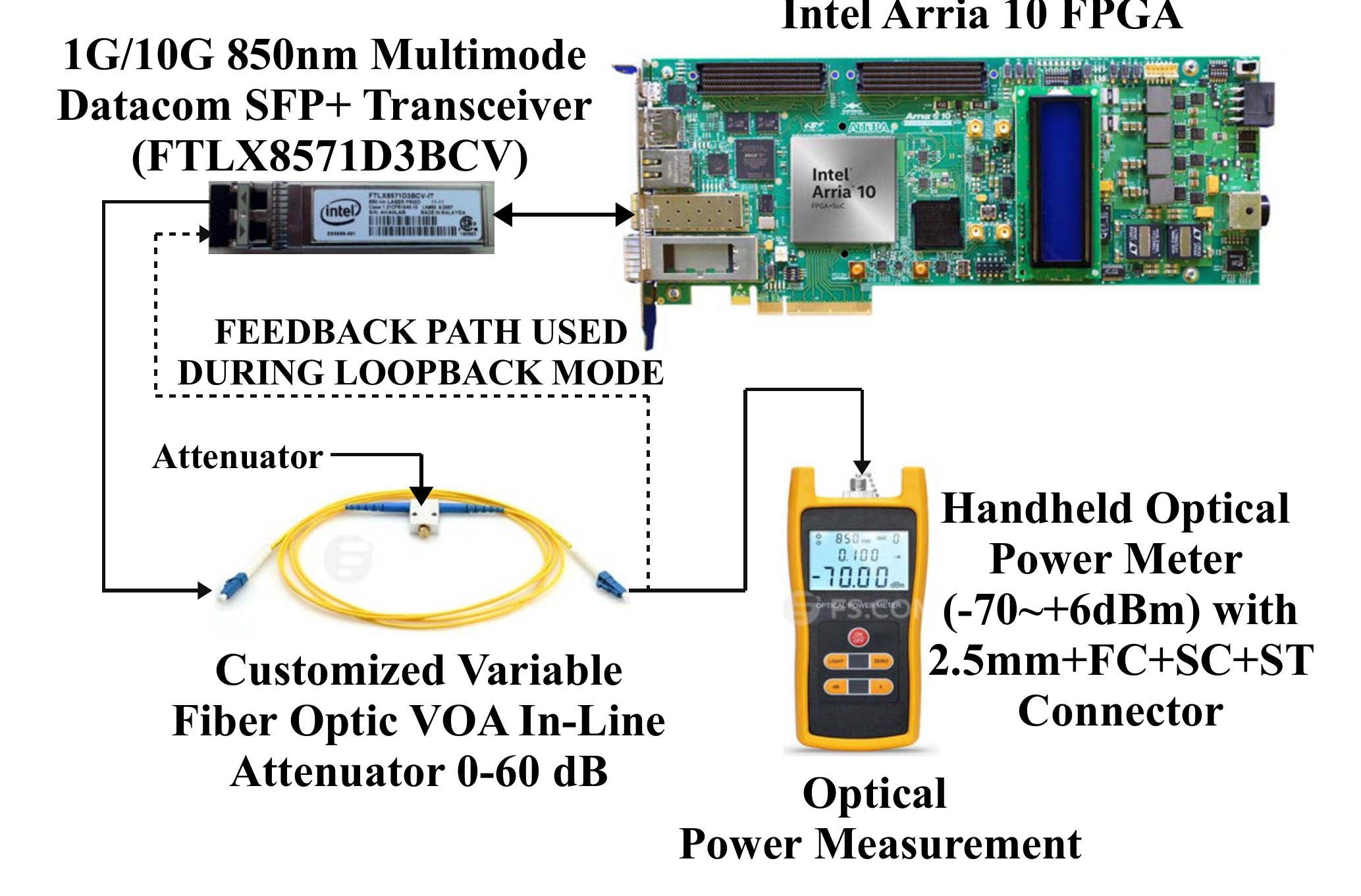}
	\caption{{GBT Measurement Setup}}
	\label{fig:setup:ber}
\end{figure}

Jitter is not the only contributing factor to bit errors; it can also be a consequence of amplitude noise. \acrfull{ber} analysis is done for quantitative measurement of signal quality. 
The Intel$ ^\circledR $ \arria development board is used for conducting the test. 
A 10G 850~nm Multimode Datacom \acrshort{sfp}+ optical transceiver is configured to operate at 4.8 Gbps, the operating line rate of the \acrshort{gbt} protocol. A customized variable fiber optic solution having in-line attenuator capability within the range of 0-60 dB is used for providing attenuation to the signal. For optical power measurement, a hand held power meter with a SC-ST connector is used. The attenuator cable adds an additional $insertion ~loss$ of $ \leq 3~dB$ to the entire test chain. Individual snapshot of the measurement setup is provided in the \fignm~\ref{fig:setup:ber}.
Interested users can use the firmware uploaded at the CERN Gitlab link \cite{jubin_link_2017} to reproduce the results in other hardware conditions. For the \acrshort{ber} measurement default transceiver configuration is used.

\acrshort{ber} is evaluated from the ratio of the number of errors received to the total number of bits transmitted. Ideally as the number of transmitted bits approaches infinity, the \acrshort{ber} becomes a perfect estimate. However, for practical tests there is a need for test procedure that allows to measure \acrshort{ber} with a high confidence level. J. Rudd~\cite{rudd2000statistical} has documented a method for reducing the test time for stressing a system, by calculating the number of bits needed to be transmitted to estimate error probability with a particular statistical confidence level. Equation~\eqref{eq:BER:CL} shows the trade-off between test time~(T) and confidence level~(CL).  

\begin{align}
\left.
\begin{tabular}{rl}
$ n $ &$ = -\dfrac{\ln(1-CL)}{BER} + \dfrac{\ln\left(\sum_{k=0}^{N} \dfrac{(n*BER)^k}{k!}\right)}{BER} $\\
$ T $&$ =n/R $
\end{tabular}\right\}
\label{eq:BER:CL}
\end{align}

where $ n $ is the total number bits transmitted,
$ N $ is the number of errors that occurred during the transmission and $ R $ is the line rate. For $ N=0 $ the solution is trivial.
\begin{figure}[!th]
	\centering
	\includegraphics[width=\linewidth]{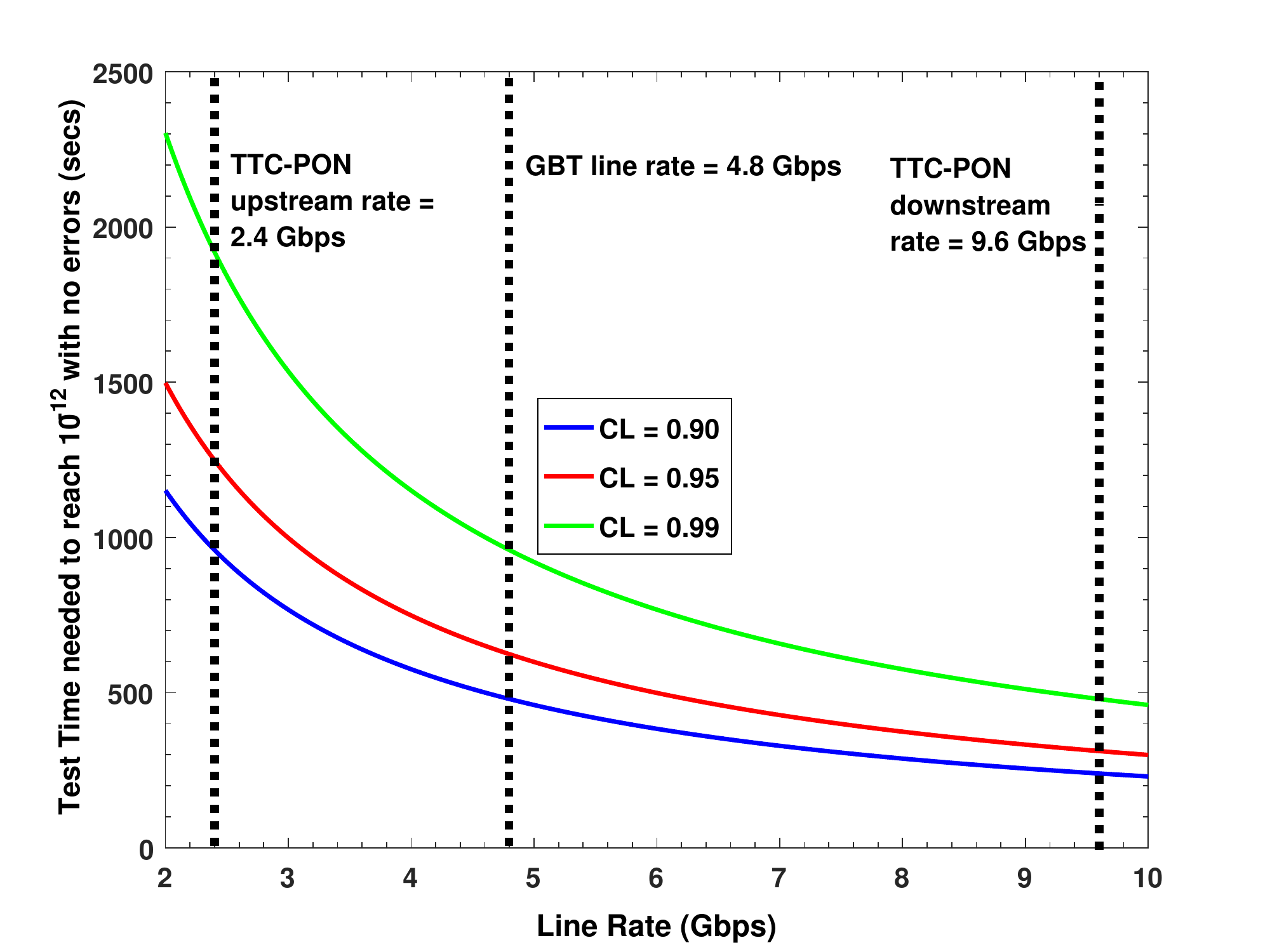}
	\caption{Time required to reach $10^{-12}$ BER vs. line rate. Showing the cases for GBT and TTC-PON.}
	\label{fig:CL}
\end{figure}
In the work of Detraz et. al. \cite{detraz2009fpga} an effort has been made to define the minimum experiment time required for \acrshort{gbt} \acrshort{ber} measurement with different level of confidence as derived from the Eq.~\eqref{eq:BER:CL}. Further the concept is extended and marked for the minimum measurement time needed for \acrshort{ttc_pon} link also, as shown in \fignm~\ref{fig:CL}.

\begin{figure}[!th]
	\centering
	\includegraphics[width=\linewidth]{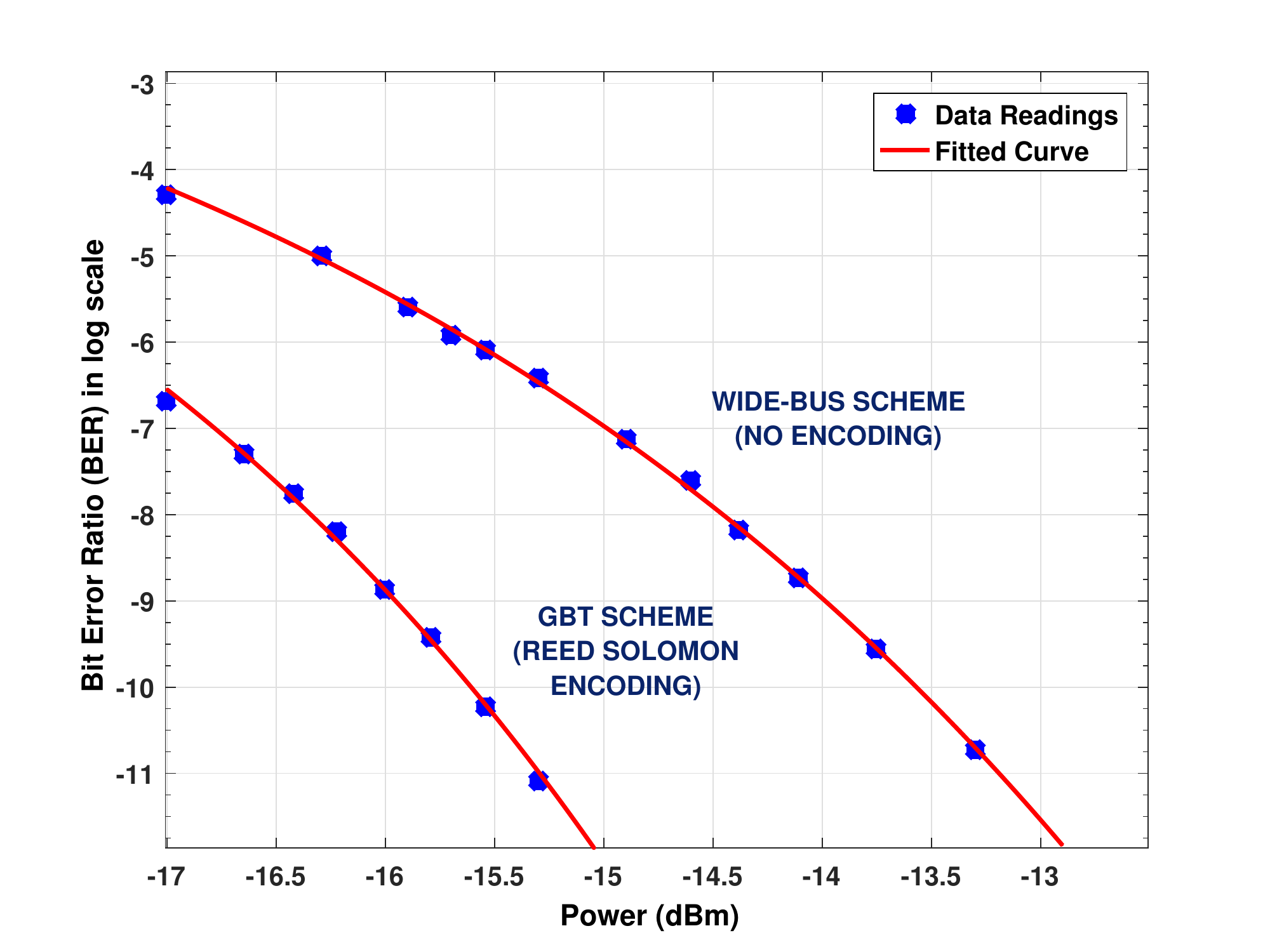}
	\caption{Showing GBT BER measurement for GBT and widebus FEC scheme }
	\label{fig:gbtber:plot}
\end{figure}

\subsubsection{BER Measurement for the GBT}
The \acrshort{gbt} \acrshort{ber} measurement for the \acrshort{gbt} encoding scheme operating in the \acrshort{gbt} mode and the Widebus mode is plotted in the \fignm~\ref{fig:gbtber:plot}. An exponential fit to the readout data is done. Below $ -17~dBm $ receiver sensitivity, due to loss of clock,  further \acrshort{ber} measurement cannot be pursued. However, the plot can be extrapolated based on standard `erfc' based nature of the curve, assuming Gaussian noise. 	
\begin{align}
&\textit{Margin of Receiver Sensitivity for targetted BER $ 10^{-12} $} \notag\\
&\textit{between both the scheme}= (15-12.9)~dBm = 2.1 ~dBm \label{eq:ber:diff}
\end{align}

The difference measured is $ 2.1 ~dBm $ as given in Eq.~\eqref{eq:ber:diff}. The result is in close agreement to the measurement conducted by Csaba Soos for \acrshort{gbt} protocol implementation implemented on  Xilinx \acrshort{fpga}~\cite{soos2008gbt}, that is around 2.5 dBm.

\begin{figure}[!th]
	\centering
	\includegraphics[width=0.95\linewidth]{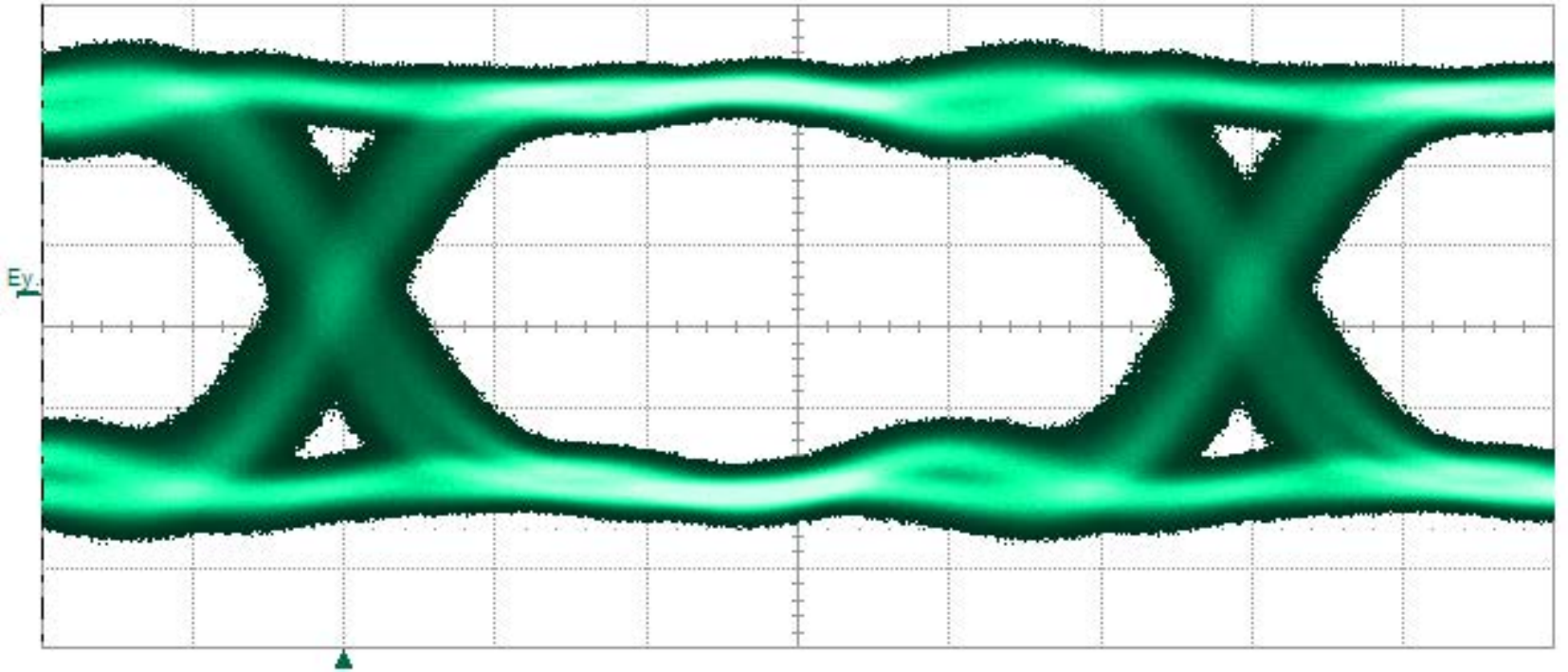}
	\caption{Showing the Eye Diagram for Tx Optical using GBT encoded data for \arria}
	\label{fig:gbt}
\end{figure}

\paragraph{The GBT link Signal Quality}
Data from the \acrshort{fpga} transceivers are transmitted using QSFP+ transceiver modules to convert the electrical signals to the optical signals for communication over a single mode fibre.
A Lecroy \acrfull{sda} oscilloscope is used for analyzing the signal quality. An eye diagram is used as an indicator to measure the quality of the optical transmission signals at the GBT line rate of 4.8 Gbps.
The signal to noise ratio of the high-speed data signal is directly indicated by the amount of eye closure
or Eye Height. For the \acrshort{gbt} transmission signal using a QSFP+ transceiver module, an \textit{eye height} of $ 406.6~mV $ and an \textit{eye width} of $ 173.3~ps $ is achieved as shown in the \fignm~\ref{fig:gbt}.

\begin{figure}[!th]
	\centering
	\begin{tabular}{c}
		\includegraphics[width=0.8\linewidth]{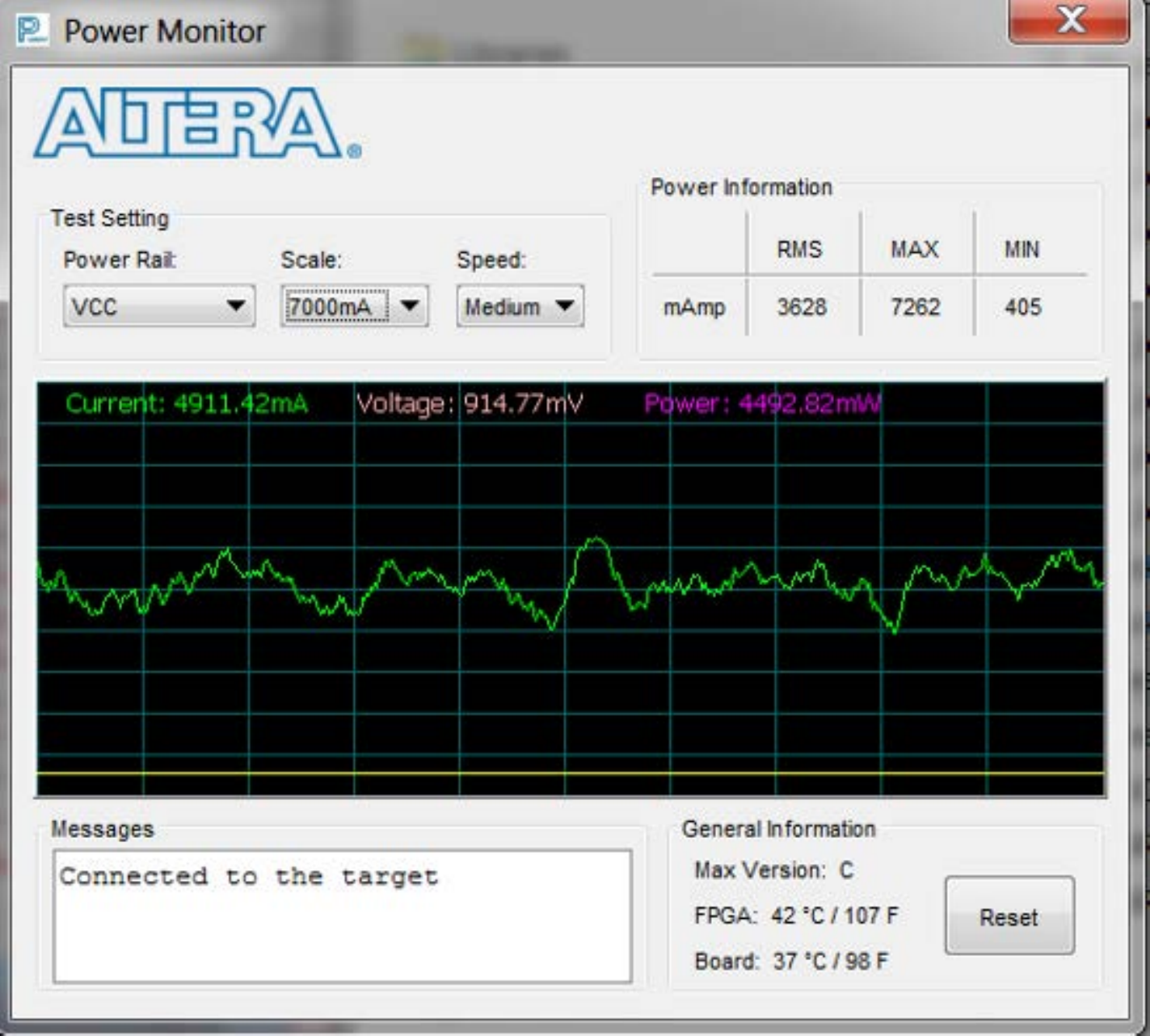} \\
		(a)\\\vspace{10ex}\\
		\includegraphics[width=0.8\linewidth]{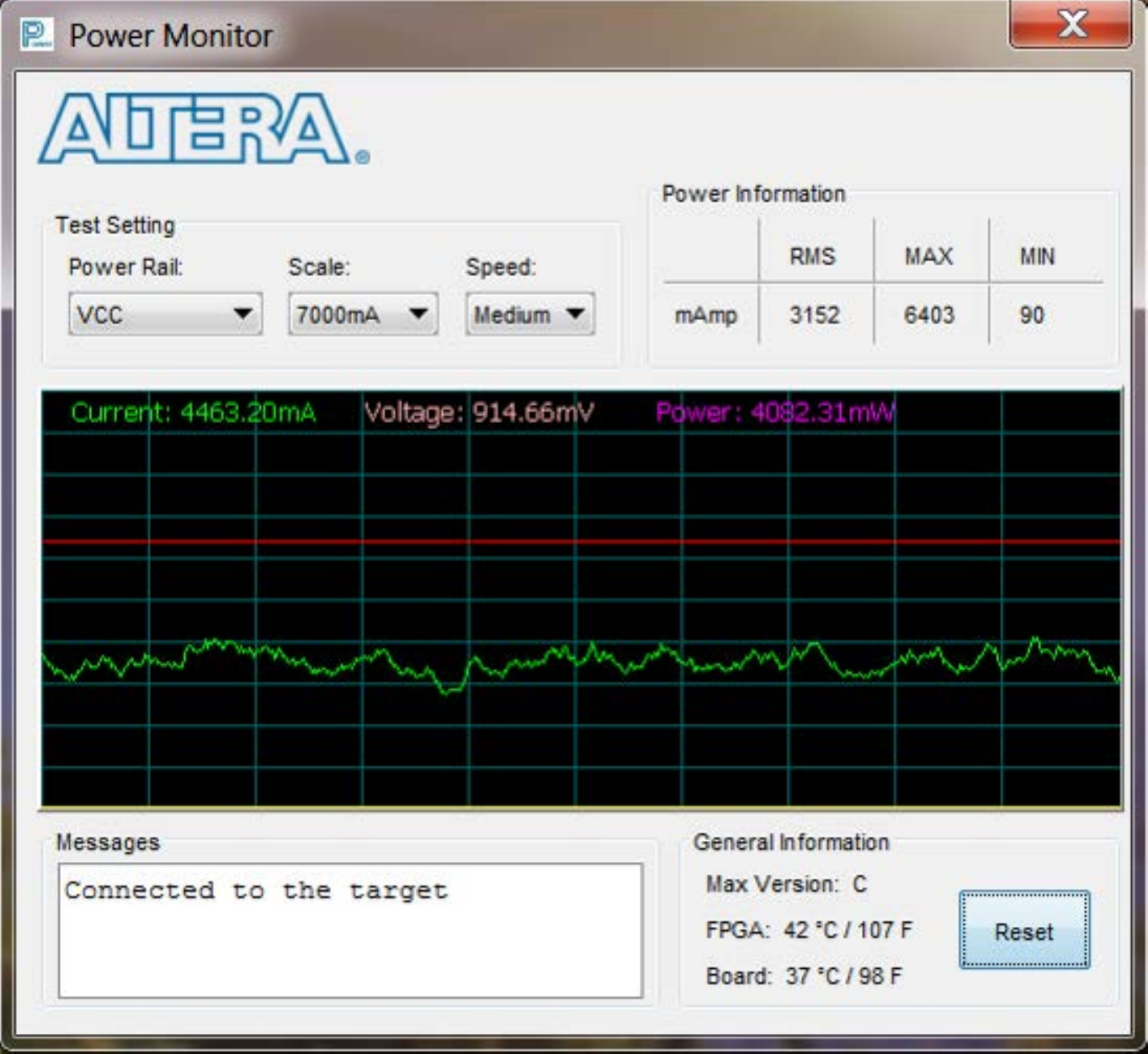}\\
		(b)
	\end{tabular}
	\caption{Showing power Monitor of GBT Firmware during data communication mode in : (a) GBT Scheme and (b) Widebus Scheme}
	\label{fig:gbt_power}
\end{figure}

\paragraph{Power Measurement}
Intel internal monitoring tools are used to register the power consumed and the temperature of the \arria \acrshort{fpga} chip during the test measurement.
\begin{table}[!th]
	\centering
	\renewcommand*{\arraystretch}{1.4}
	\caption{Power consumed by a single GBT link for different encoding scheme}
	\label{table:gbt_power}
	\resizebox{\linewidth}{!}{
		\begin{tabular}{c|c|c|c|c|c}
			\hline
			\textbf{Encoding}  & \textbf{Board Temp.}  & \textbf{FPGA Temp.} & \textbf{RMS Current} ($ I_{RMS} $) & \textbf{Voltage} ($ V $)  & \textbf{Power} ($ P=V\times I_{RMS} $) \\
			\textbf{Scheme} &$^\circ~C$ & $^\circ~C$ &  (mA) & (mV) & (mW)\\\hline
			\textbf{GBT} & 37 & 42 & 4911.42 & 914.77 & 4492.82\\
			\textbf{Widebus} & 37 & 42 & 4463.20 & 914.66 & 4082.31\\\hline
	\end{tabular}}
\end{table}
\fignm~\ref{fig:gbt_power} shows the power variation plot as monitored using the tool and \tablenm~\ref{table:gbt_power} summarizes the results collected when a single \acrshort{gbt} link under the different encoding scheme consumes different power.

In the \acrshort{cru} project, the link connecting the radiation hard component to the non-radiation hard component, is based on the \acrshort{gbt} link technology operating at 4.8 Gbit/s using a 850~nm multimode optical fibre. The link is connected between the versatile link transceiver to a \acrfull{mpo} optical connector at the \acrshort{cru} PCIe40 \acrshort{daq} Engine using an optical fibre splitter. A study has already been conducted by Schwemmer et. al. \cite{schwemmer2014evaluation} to note the performance on 400~m long OM3 and OM4 cables. So, further test on optical cable characterization is not pursued.


\subsubsection{BER Measurement for the TTC-PON}
\begin{figure}[!th]
	\centering
	\includegraphics[trim={150 270 140 10},clip,width=0.65\linewidth]{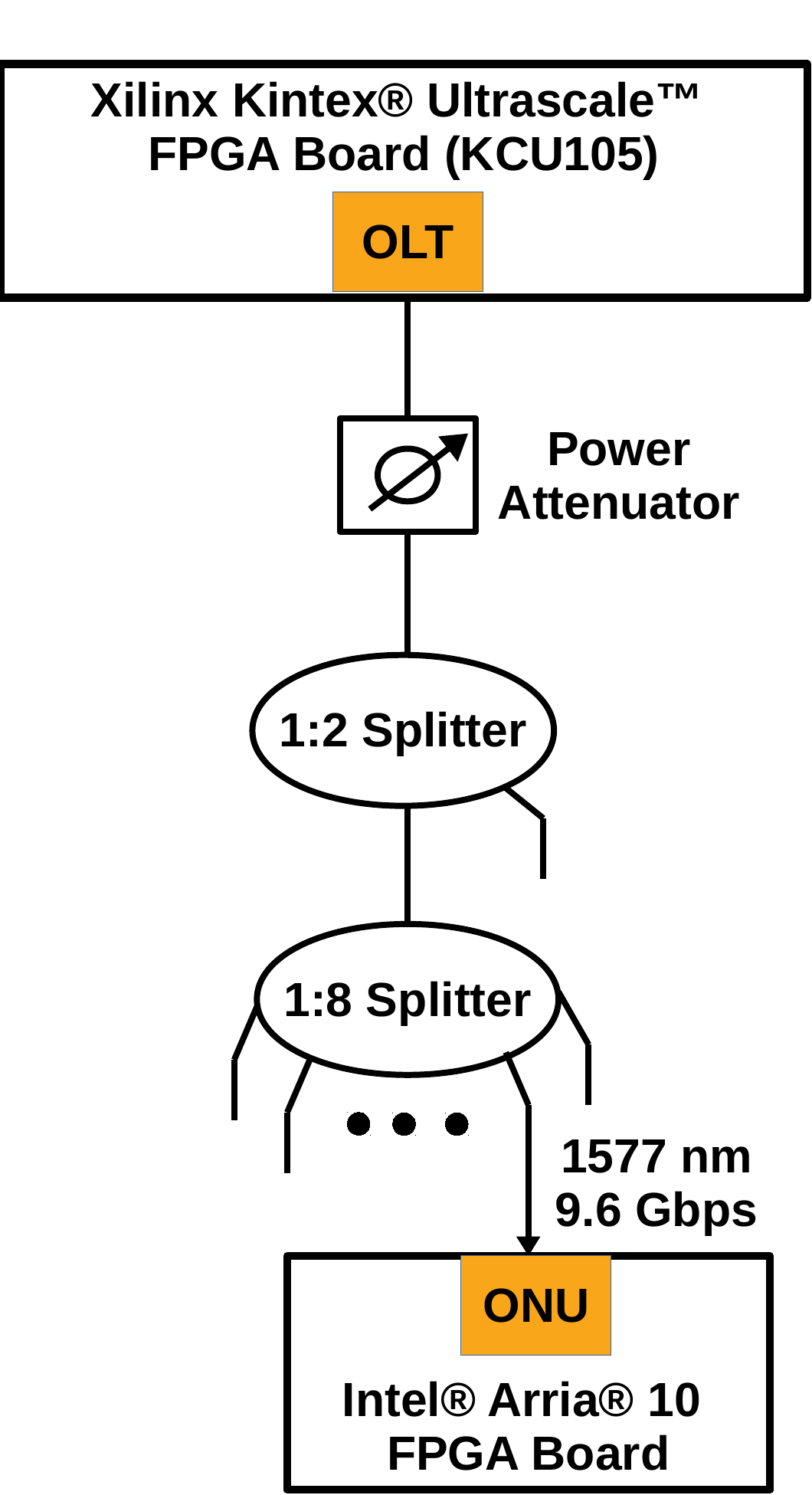}
	\caption{The TTC PON signal quality measurement setup.}
	\label{fig:ttc_pon_setup}
\end{figure}

The test setup used for TTC-PON BER measurement is different from the GBT BER measurement.
The test setup involves an optical connection from the Kintex$^ \circledR  $ ~Ultrascale$ ^{TM} $ FPGA having a \acrfull{olt} module to the Intel$^ \circledR  $  \arria FPGA having a \acrfull{onu} module. For emulating the real experiment conditions an inline power attenuator and an optical splitter is used. Details of the experimental setup is illustrated graphically in the \fignm~\ref{fig:ttc_pon_setup}.
The transmission part on the \acrshort{onu}-\acrshort{sfp} on the Intel \acrshort{fpga} board was disabled for the upstream path as it was not implemented during the pre-validation measurement test. The ONU just acted as a receiver. 
A \acrfull{prbs} generator was incorporate in the \acrshort{olt} Kintex$^ \circledR  $ ~Ultrascale$ ^{TM} $ design side while a \acrshort{prbs} checker was used in the \acrshort{onu} \arria design side in order to perform the test.
Attenuation was changed in the range of 10.0~dB~:~1.0~dB~:~20.0~dB. For each attenuation value, the average optical received power at the ONU level was measured using the JDSU OLP-87 \acrshort{pon} Power Meter. The results are tabulated in the \tablenm~\ref{table:xcvr_attn}. Transmitted power of the \acrshort{olt} is +3.67 dBm.  For each change in the attenuation value, $  5\times 10^{11} $ bits are transmitted to measure the quantity of errors.

\begin{figure}[!th]
	\centering 
	\includegraphics[width=\linewidth]{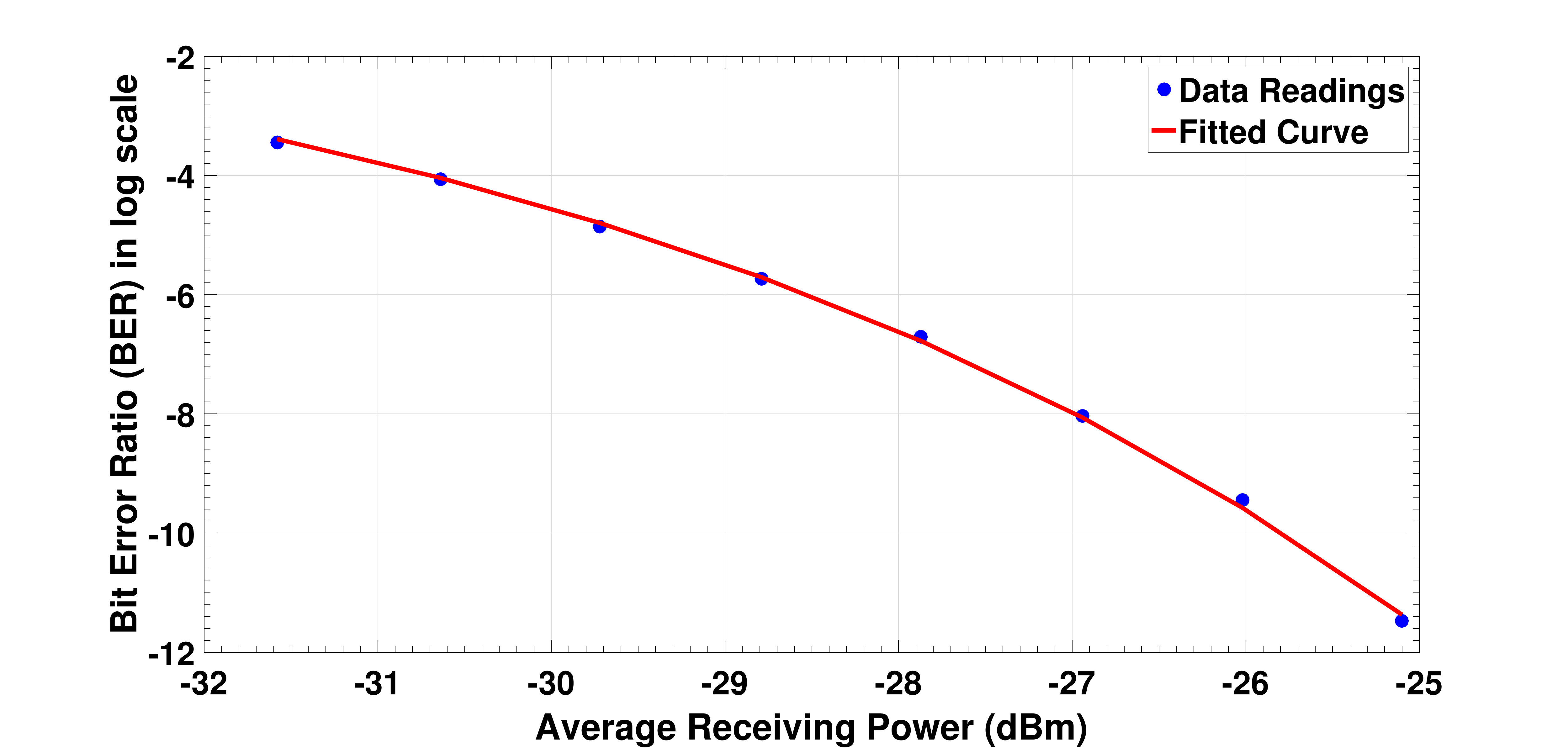}
	\caption{Downstream Bit Error Rate (BER) from the OLT to the  ONU}
	\label{fig:ber}
\end{figure}

\begin{table}[!th]
	\centering
	\caption{Variation of the received power with change in the attenuation}
	\label{table:xcvr_attn}
	\renewcommand{\arraystretch}{1.2}
	\begin{tabular}{cc}
		\hline
		\textbf{Attenuation (dB)} & \textbf{Received Power (dBm)}\\\hline
		20 & -31.58\\
		19 & -30.64\\
		18 & -29.72\\
		17 & -28.79\\
		16 & -27.87\\
		15 & -26.94\\
		14 & -26.02\\
		13 & -25.10\\
		12 & -24.19\\
		11 & -23.25\\
		10 & -22.32\\
		0 & -13.13\\\hline
	\end{tabular}
\end{table}

\begin{figure}[!th]
	\centering 
	\includegraphics[width=\linewidth]{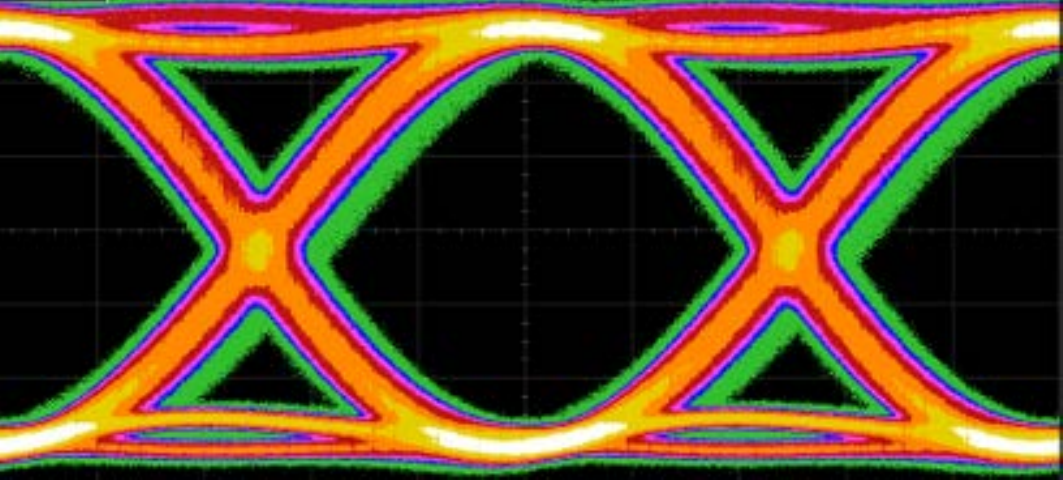}
	\caption{TTC PON signal quality analysis in an Agilent Scope by plotting an eye diagram}
	\label{fig:KCU105_LTF7222_TX_EYE}
\end{figure}

\begin{figure}[!th]
	\centering 
	\includegraphics[width=\linewidth]{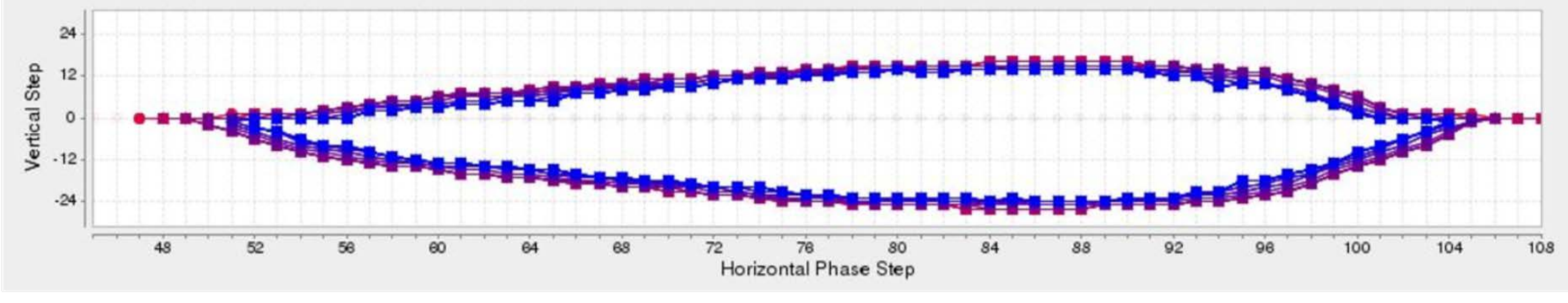}
	\caption{TTC PON signal quality monitoring using Transceiver Toolkit (TTK) in serial loopback mode.}
	\label{fig:ttc_pon_serial_loopback}
\end{figure}

\paragraph{The TTC PON \arria Signal Quality}
\fignm~\ref{fig:KCU105_LTF7222_TX_EYE} shows the transmission quality of the optical signal of the \acrshort{ttc_pon} from the \acrshort{ctp} prototype board, that uses a Xilinx Ultrascale \acrshort{fpga}. The \fignm~\ref{fig:ttc_pon_serial_loopback} shows the receiving signal quality of the \acrshort{ttc_pon} as monitored within the \arria FPGA using a transceiver toolkit (TTK). The Eye Width to the Eye height ratio of 53/39 is registered in the TTK for the test bits of 1.1E12 using the PRBS31 stress pattern.

\subsection{Transceiver Optimization}
Transceiver parameter tuning plays a significant role to reduce BER.
A test procedure is developed to tune the high-speed link using the signal conditioning circuitry provided in the \arria transceivers. 
Quartus v15.1 transceiver testing toolkit \cite{altera2012xcvr} is used to monitor the signal characteristics.
Several articles are mentioned in the Altera (now Intel) literature the need for proper optimization of transceiver for a maximum performance   \cite{altera2015hps} \cite{altera2012xcvr} \cite{alteraBlake} \cite{emphasis2010altera}. All those articles are dedicated to the old generation FPGAs like Stratix IV, Cyclone and others, hence a study was necessary to have a first-hand result of the effect of the transceiver tuning on the \arria FPGAs \cite{native_phy_arria}.  For the transceiver tests, the line rate of the \acrshort{gbt} that is 4.8~Gbps is used in an optical loopback mode. The majority of the links used in the \acrshort{cru} are of \acrshort{gbt} standard hence the \acrshort{gbt} link rate is selected to carry out the test.

The transceiver optimization by testing for all the combinations would be an inefficient approach as the time required would be very long. A short calculation is given to show the exact measurement time needed to scan for all the configurations. Empirically the reading time for each test configurations is found to be of 10 secs. With the allowed modifiable transceiver properties in \arria, the configuration range possible for each parameter is listed in the \tablenm~\ref{table:default_values}. 
The total number of tuples of the configuration cases possible is a pure product of all the test cases, which is given by 32 x 63 x 63 x 31 x 15 x 16 x 8 or 7,559,516,160 (approx 7.5 billion) test cases. 
The total time needed to execute all the configurations is 2397~years (=~7,559,516,160 x 10~secs =~20,998,656~hrs =~874,944~days).

\begin{table}
	\centering
	\caption{The default range of configuration parameters}
	\label{table:default_values}
	\begin{tabular}{lrcrc}
		\hline 
		\textbf{Parameters}  & \multicolumn{3}{c}{\textbf{Range}} & \textbf{No. of cases}\tabularnewline
		\hline 
		\hline 
		V{\tiny{}OD}  & 00 &  \textendash{}  & 31 & 32\tabularnewline
		Pre-emphasis 1st post-tap  & - 31 &  \textendash{}  & 31 & 63\tabularnewline
		Pre-emphasis 1st pre-tap  & - 31 &  \textendash{}  & 31 & 63\tabularnewline
		Pre-emphasis 2nd post-tap  & - 15 &  \textendash{}  & 15 & 31\tabularnewline
		Pre-emphasis 1st post-tap  & - 07 &  \textendash{}  & 07 & 15\tabularnewline
		Equalization  & 00 & \textendash{}  & 15 & 16\tabularnewline
		VGA  & 00 &  \textendash{}  & 07 & 08\tabularnewline
		\hline 
	\end{tabular}
\end{table}

\begin{figure*}[!th]
	\centering
	\includegraphics[trim={20 2650 10 50},clip,width=\linewidth]{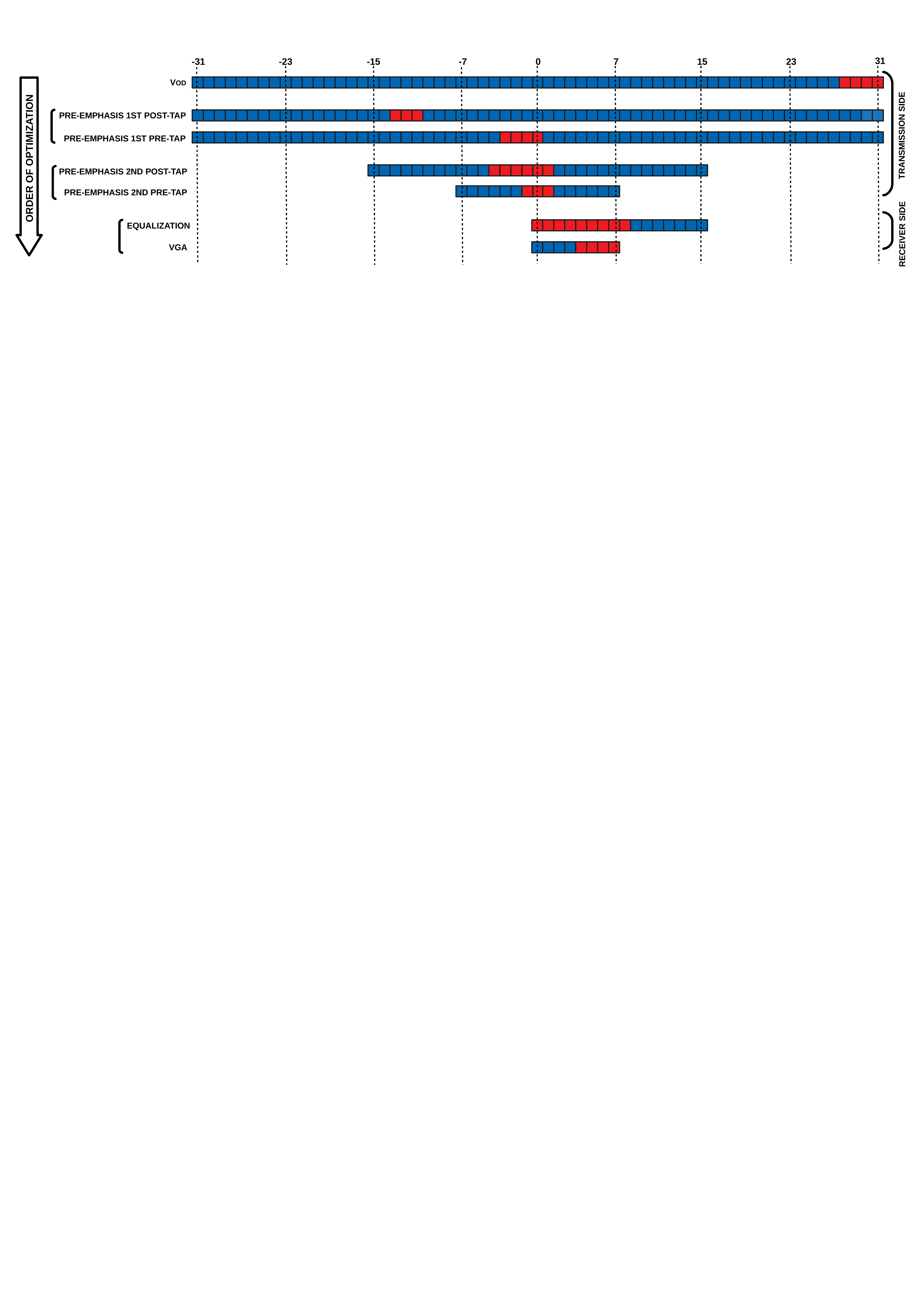}
	\caption{Choice of the optimized values from the complete range of configuration parameters}
	\label{fig:sub_optimal_parameters}
\end{figure*}

\begin{figure*}[!th]
	\centering
	\includegraphics[trim={290 70 290 50},clip,width=0.75\linewidth]{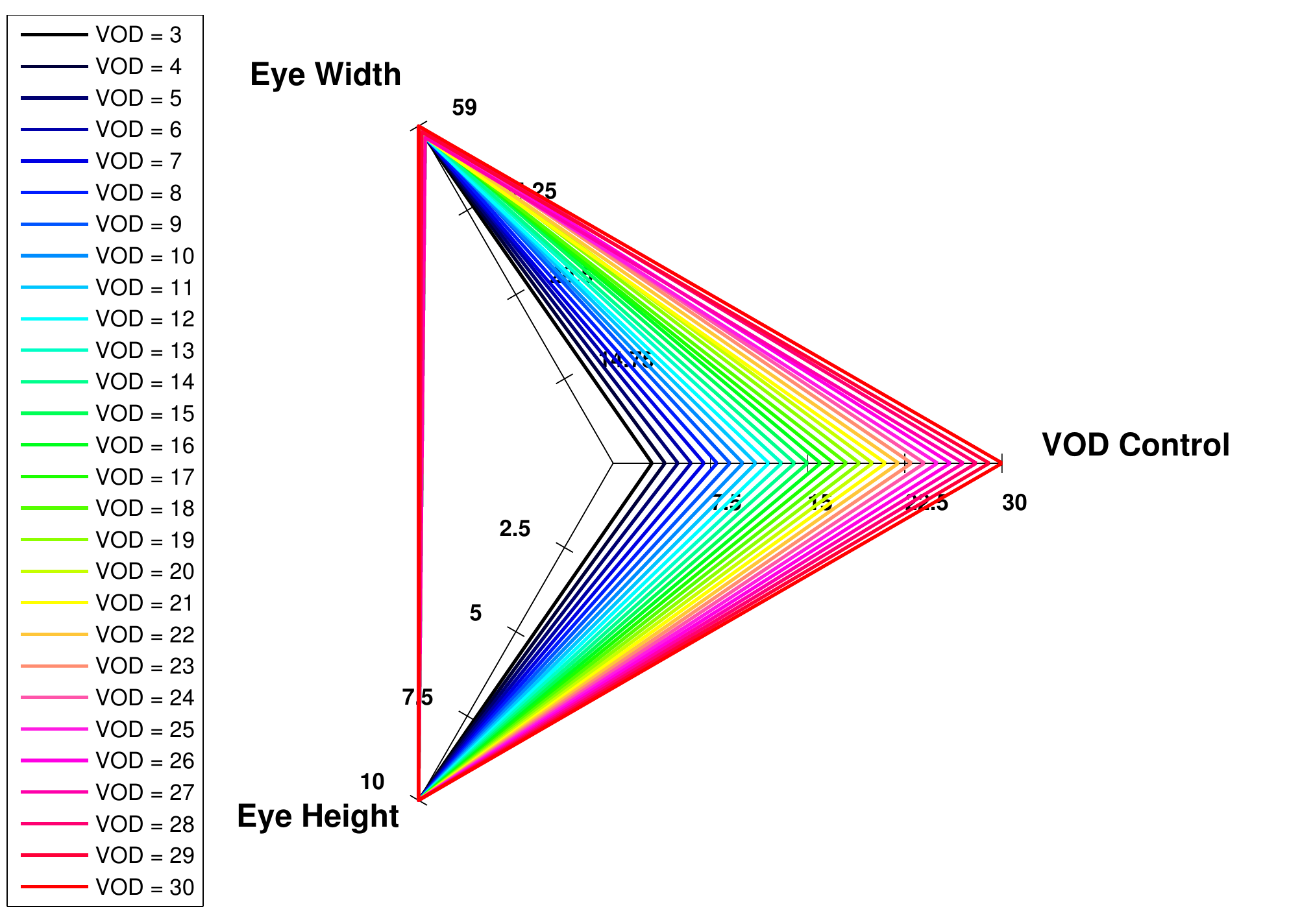}
	\caption{{Transceiver parameter tuning plot with VOD variation}}
	\label{fig:xcvr:vod}
\end{figure*}

Instead of spending huge computational time to look for an optimal solution, a good enough workaround is to find a potential suboptimal solution.
Individual parameter configuration range are scanned by keeping all other parameters fixed at Intel default values, and obtained a range of optimized values determined from the eye-width to the eye-height ratio in the EyeQ signal monitoring tool. 
The linear nature of the configuration parameters causes the evaluated optimized values to appear in contiguous subsequence as depicted in the \fignm~\ref{fig:sub_optimal_parameters}.
Out of the multiple parameters, the variation of only V\textsc{OD} parameter against the eye height and the eye width using a spider chart is plotted in the \fignm~\ref{fig:xcvr:vod} to illustrate the procedure.

The parameters can be grouped to optimize separately without having any notable interference effect on the adjacent parameters. Like \{VOD\}, \{Pre-Emphasis 1st Post Tap, Pre-Emphasis 1st Pre Tap\}, \{Pre-Emphasis 2nd Post Tap, Pre-Emphasis 2nd Pre Tap\} and \{VGA, EQUALIZATION\} can be grouped separately. However, for the fast approximation, the order of optimization is kept same as in the order mentioned. The pictorial representation is illustrated in the \fignm~\ref{fig:sub_optimal_parameters}. Users can change the order for further optimization. By this method, the time taken reduced significantly. The total number of configuration cases comes down to 70  ( = 4 + (3 x 4) + (6 x 3) + (9 x 4)).
The total time needed is of 11.66 mins (= 70 x 10~secs = 700~secs).  The major reason to opt for the quick estimation is to characterize for more than 24 links per CRU board and repeat it for over 100 CRU boards in a short period of time where the result of one board cannot be applied to the other board. Hence, this \textit{heuristic approach} is developed.

\begin{table}
	\centering
	\caption{The optimized range of individual configuration parameters}
	\label{table:best_values}
	\begin{tabular}{lrcrc}
		\hline 
		\textbf{Parameters} & \multicolumn{3}{c}{\textbf{Range}} & \textbf{No. of cases}\tabularnewline
		\hline 
		\hline 
		V{\tiny OD} & 28 & \textendash{} & 31 & 4\tabularnewline
		Pre-emphasis 1st post-tap & - 13 & \textendash{} & - 11 & 3\tabularnewline
		Pre-emphasis 1st pre-tap & - 03 & \textendash{} & 00 & 4\tabularnewline
		Pre-emphasis 2nd post-tap & - 04 & \textendash{} & 01 & 6\tabularnewline
		Pre-emphasis 1st post-tap & - 01 & \textendash{} & 01 & 3\tabularnewline
		Equalization & 00 & \textendash{} & 08 & 9\tabularnewline
		VGA & 04 & \textendash{} & 07 & 4\tabularnewline
		\hline 
	\end{tabular}
\end{table}

\begin{figure*}[!th]
	\centering
	\includegraphics[trim={200 100 200 50},clip,width=0.9\linewidth]{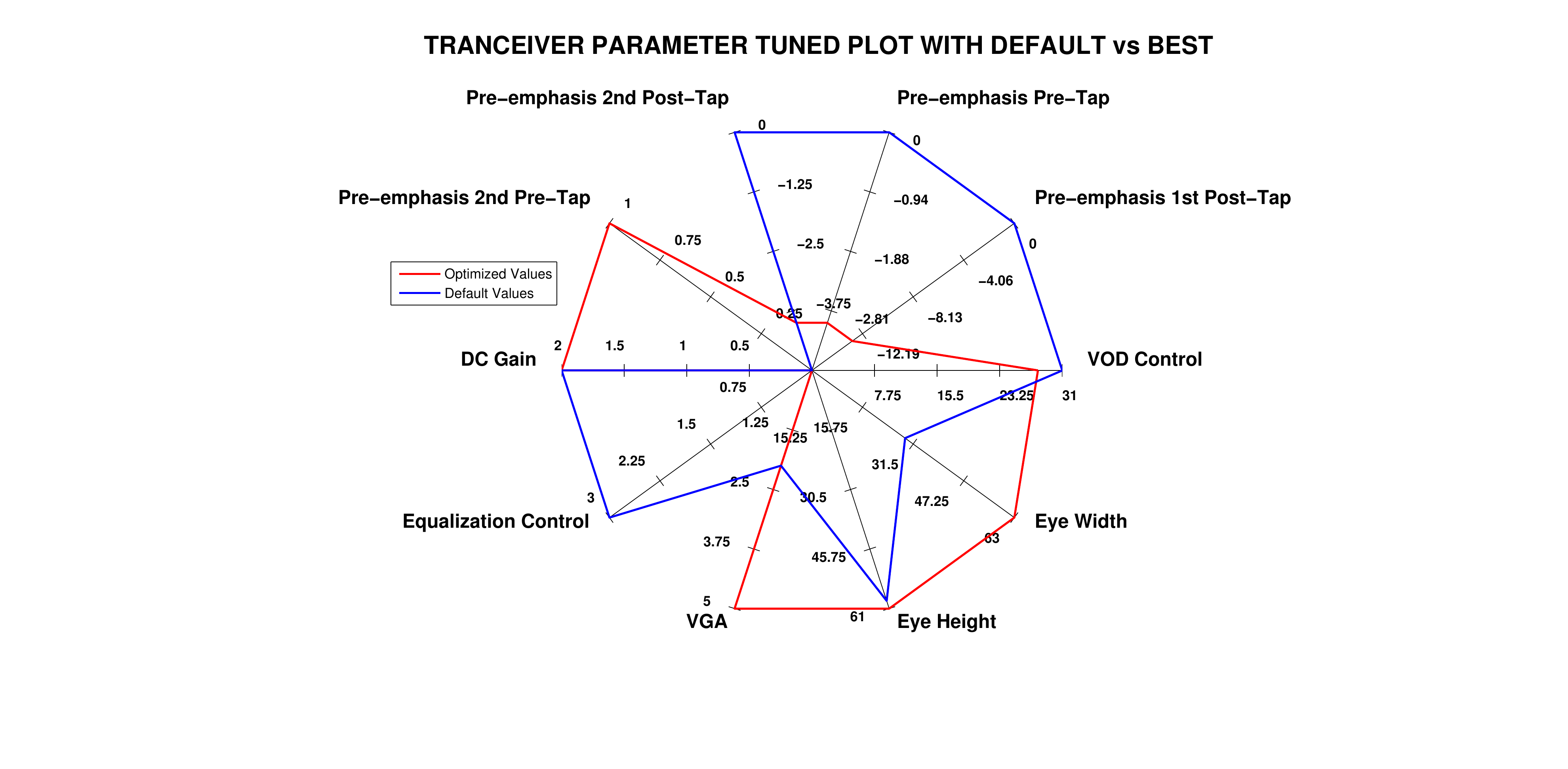}
	\caption{{Transceiver parameter tuned showing values for the default settings vs the best settings}}
	\label{fig:xcvr:tuned_parameter}
\end{figure*}

\begin{figure*}[!th]
	\begin{tabular}{c}
		\includegraphics[width=\linewidth]{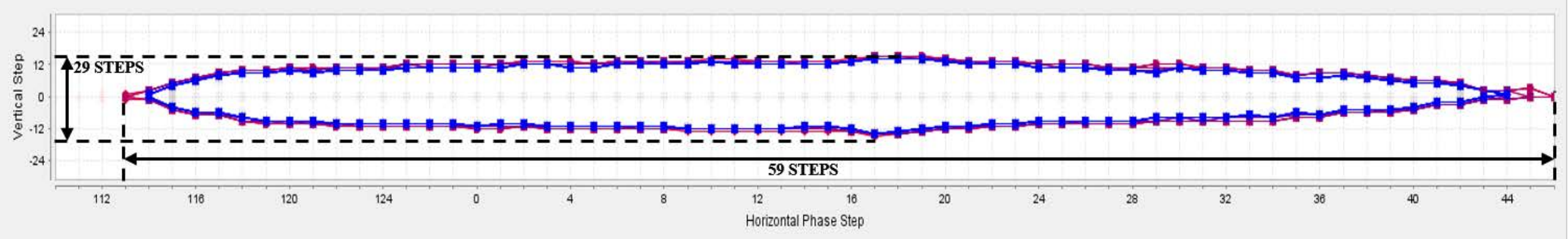}\\
		(a)\\\\
		\includegraphics[width=\linewidth]{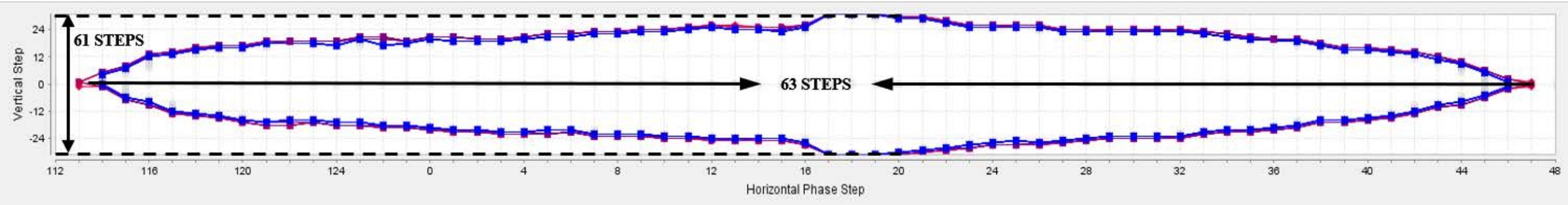}\\
		(b)\\
	\end{tabular}
	\caption{Eye Diagram (a) Default parameter settings; (b) Optimized parameter settings}
	\label{fig:eye:diagram}
\end{figure*}

After the parameters are optimized, the values for the default and the best conditions are shown in the \fignm~\ref{fig:xcvr:tuned_parameter}.
For the entire test, PRBS31 is used to stress the system as it shows the maximum bit transitions among the other available patterns. 
The eye diagram against the default parameter and the optimized parameter are shown in the \fignm~\ref{fig:eye:diagram}. 
The different parameter configurations of the transceiver can share the same eye diagram values or the performance metrics. The set of those values or the configuration parameter tuples are referred as the \textit{solution space}. The obtained best configuration parameters of the device under test is tabulated in the \tablenm~\ref{table:optimized_values}. The optimized configuration values are highly sensitive and depend on the FPGA process technology, the system temperature and the optical transceiver used. Hence, even for a minor hardware modification, the configuration parameter values need to be re-evaluated. 

\begin{table}
	\centering
	\caption{The optimized value of configuration parameters}
	\label{table:optimized_values}
	\begin{tabular}{lr}
		\hline
		\textbf{Parameters} & \textbf{Optimized Value} \\
		\hline\hline
		V{\tiny OD} & 28\\
		Pre-emphasis 1st post-tap & -13\\
		Pre-emphasis 1st pre-tap  & -03\\
		Pre-emphasis 2nd post-tap & -04\\
		Pre-emphasis 1st post-tap & 01\\
		Equalization & 00\\
		VGA          & 05\\\hline
	\end{tabular}
\end{table}

%
%
%
%
%

\section{Discussions}
\label{sec:discussion}
The behaviour of all the composite elements in the \acrshort{ttc_pon} and \acrshort{gbt} bridge connection is analyzed for compatibility regarding interconnection and interoperability.
A detailed characterization test on the integrated prototype design is conducted to check for any unanticipated design faults before the final commissioning in CRU firmware. Four performance measuring metrics are used in the characterization: (1) Latency; (2) Jitter; (3) BER and (4) Transceiver parameter settings. 
The testing phase of the firmware has passed through several iterations of \acrfull{pfr} for the reliability test of the communication bridge.
During the entire study, large sets of empirical results are collected for analysis. The analysis of the statistics confirms that the TTC communication bridge connection with its present design configuration showed complete deterministic behaviour with no points of uncertainty over the several rounds of \acrshort{pfr}.

The results meet the specified trigger and timing communication standards hence no further compensatory measures are needed. 
To avoid any unprecedented failure during the data taking; a set of monitoring logic is integrated along with the CRU firmware logic core to register the macroscopic behaviour of the system, as a preventive measure, as mentioned in Section~\ref{sec:design_resilence}.
The intrinsic details related to the \acrshort{cru} firmware designs are available at the \acrshort{alice}-\acrshort{cern} \acrshort{cru} TWiki page \cite{alice_cru_twiki}.

\section{Summary and Outlook}
\label{sec:summary}

We have carried out a detailed study of the trigger and timing distributions using the TTC-PON and GBT bridge connection in ALICE. 
The study is carried out using the CRU development boards for rapid dissemination of performance metrics. The results show that the TTC-PON and GBT can work in synergy to communicate successfully the timing and trigger information and can effectively be deployed. 
The study confirmed that the system behaviour is completely deterministic 
with multiple rounds of \acrshort{pfr}. 
The \acrshort{fpga} used in the \acrshort{cru} board is  $ 20~nm $ Intel$ ^\circledR $ \arria. 
The CRU firmware logic uses static placement configuration, hence the stress points remain fixed over the operation runtime. 
Future scope is to do a reliability study by accelerated stress scenario to mimic the effect of degradation in the timing circuits and wearability in the logic/memory cells~\cite{Stott2011}. These would identify the stress hotspots and allow us to overcome the system faults by applying the mitigation solutions accordingly. 
Another study that is equally important is to have a data flow analysis of the spatiotemporal behaviour of the data traffic~\cite{wang2002capturing} for each sub-detector system to arrange and reallocate the~\acrshort{cru} peripheral logic resources in an optimized manner.

\section{Acknowledgement}
We would like to express our sincere gratitude to Jean-Pierre Cahcemeiche and the team 
members at  CPPM, Marseille for providing constant support in the PCIe40 board usage while conducting the tests.
We would also like to thank Marian Krivda and Roman Lietava from the \acrshort{alice} Trigger Group for their valuable suggestions.

\section{References}
\bibliography{reference}

\begin{thebibliography}{10}
\expandafter\ifx\csname url\endcsname\relax
  \def\url#1{\texttt{#1}}\fi
\expandafter\ifx\csname urlprefix\endcsname\relax\def\urlprefix{URL }\fi
\expandafter\ifx\csname href\endcsname\relax
  \def\href#1#2{#2} \def\path#1{#1}\fi

\bibitem{abelav2014aliceUpgrade}
{B Abelev et. al. and the ALICE Collaboration},
  \href{http://iopscience.iop.org/article/10.1088/0954-3899/41/8/087001}{{Upgrade
  of the ALICE Experiment: Letter Of Intent}}, Journal of Physics G: Nuclear
  and Particle Physics 41~(8) (2014) 087001.

\bibitem{alex2013readout}
P.~Antonioli, A.~Kluge, W.~Riegler, {for the ALICE Collaboration},
  \href{http://cds.cern.ch/record/1603472}{Upgrade of the ALICE Readout \&
  Trigger System}, CERN Technical Design Report CERN-LHCC-2013-019,
  ALICE-TDR-015.

\bibitem{cachemiche2016pcie}
J.~Cachemiche, P.~Duval, F.~Hachon, R.~Le~Gac, F.~R{\'e}thor{\'e},
  \href{http://iopscience.iop.org/article/10.1088/1748-0221/11/02/P02013/meta}{The
  PCIe-based readout system for the LHCb experiment}, Journal of
  Instrumentation 11~(02) (2016) P02013.

\bibitem{alicePerf2014}
{The ALICE Collaboration},
  \href{http://www.worldscientific.com/doi/abs/10.1142/S0217751X14300440}{Performance
  of the ALICE experiment at the CERN LHC}, International Journal of Modern
  Physics A 29~(24) (2014) 1430044.
\newblock \href {http://dx.doi.org/10.1142/S0217751X14300440}
  {\path{doi:10.1142/S0217751X14300440}}.

\bibitem{alice2014hlt}
{ALICE Collaboration},
  \href{https://compeng.uni-frankfurt.de/wiki/HLT/images/7/78/Alice_tdr_trigger_hlt_daq.pdf}{ALICE
  trigger data-acquisition high-level trigger and control system}, Technical
  Design Report ALICE.

\bibitem{costa2017readout}
F.~Costa, A.~Kluge, P.~V. Vyvre, A.~Collaboration,
  \href{http://stacks.iop.org/1742-6596/898/i=3/a=032011}{The detector read-out
  in ALICE during Run 3 and 4}, Journal of Physics: Conference Series 898~(3).

\bibitem{Buncic2015online}
P.~Buncic, M.~Krzewicki, P.~Vande~Vyvre, {for the ALICE Collaboration},
  \href{https://cds.cern.ch/record/2011297?ln=en}{Upgrade of the Online-Offline
  Computing System}, CERN Technical Design Report CERN-LHCC-2015-006 ;
  ALICE-TDR-019.

\bibitem{alice_upgrade_16}
A.~Kluge,
  \href{https://indico.cern.ch/event/468486/contributions/1144350/attachments/1239193/1821618/20160307ACES.pdf}{ALICE
  upgrade in LS2}, {ACECS 2016-Fifth Common ATLAS CMS Electronics Workshop for
  LHC Upgrades}.

\bibitem{Kyung2011jitter}
X.~C. C.~Y. {Kyung Suk (Dan) Oh},
  \href{https://www.amazon.in/High-Speed-Signaling-Modeling-Budgeting-Semiconductor/dp/0132826917}{High-Speed
  Signaling: Jitter Modeling, Analysis, and Budgeting}, 1st Edition, Prentice
  Hall Modern Semiconductor Design Series, 2011.

\bibitem{gbt09moreira}
P.~Moreira, R.~Ballabriga, S.~Baron, S.~Bonacini, O.~Cobanoglu, F.~Faccio,
  T.~Fedorov, R.~Francisco, P.~Gui, P.~Hartin, et~al.,
  \href{https://cds.cern.ch/record/1235836}{The GBT project}, in: {Proceedings
  of the Topical Workshop on Electronics for Particle Physics}, 2009, pp.
  342--346.

\bibitem{gbt_baron}
S.~Baron, J.~Cachemiche, F.~Marin, P.~Moreira, C.~Soos,
  \href{https://cds.cern.ch/record/1236361?ln=en}{Implementing the GBT data
  transmission protocol in FPGAs}, in: Topical Workshop on Electronics for
  Particle Physics (TWEPP), 2009.

\bibitem{amaral2009versatile}
L.~Amaral, S.~Dris, A.~Gerardin, T.~Huffman, C.~Issever, A.~J. Pacheco,
  M.~Jones, S.~Kwan, S.-C. Lee, Z.~Liang, et~al.,
  \href{http://iopscience.iop.org/article/10.1088/1748-0221/4/12/P12003/meta}{The
  versatile link, a common project for super-LHC}, Journal of Instrumentation
  4~(12) (2009) P12003.

\bibitem{kolotouros2015ttc}
D.-M. Kolotouros, S.~Baron, C.~Soos, F.~Vasey,
  \href{http://iopscience.iop.org/article/10.1088/1748-0221/10/04/C04001/pdf}{A
  TTC upgrade proposal using bidirectional 10G-PON FTTH technology}, Journal of
  Instrumentation 10~(04) (2015) C04001.

\bibitem{verissimo1997role}
P.~Ver{\'\i}ssimo, \href{http://ieeexplore.ieee.org/document/644744/}{On the
  role of time in distributed systems}, in: Distributed Computing Systems,
  Proceedings of the Sixth IEEE Computer Society Workshop on Future Trends of,
  IEEE, 1997, pp. 316--321.

\bibitem{marin2015gbt}
M.~B. Marin, S.~Baron, S.~Feger, P.~Leitao, E.~Lupu, C.~Soos, P.~Vichoudis,
  K.~Wyllie,
  \href{http://iopscience.iop.org/article/10.1088/1748-0221/10/03/C03021}{The
  GBT-FPGA core: features and challenges}, Journal of Instrumentation 10~(03)
  (2015) C03021.

\bibitem{baron2009implementing}
S.~Baron, J.~Cachemiche, F.~Marin, P.~Moreira, C.~Soos,
  \href{http://hal.in2p3.fr/in2p3-00468912/document}{Implementing the GBT data
  transmission protocol in FPGAs}, in: TWEPP-09 Topical Workshop on Electronics
  for Particle Physics, 2009, pp. 631--635.

\bibitem{mitra2016gbt}
J.~Mitra, S.~A. Khan, M.~B. Marin, J.~P. Cachemiche, E.~David, F.~Hachon,
  F.~Rethore, T.~Kiss, S.~Baron, A.~Kluge, et~al.,
  \href{http://iopscience.iop.org/article/10.1088/1748-0221/11/03/C03039}{GBT
  link testing and performance measurement on PCIe40 and AMC40 custom design
  FPGA boards}, Journal of Instrumentation 11~(03) (2016) C03039.

\bibitem{pap09pon}
S.~Papadopoulos, I.~Darwazeh, I.~Papakonstantinou, J.~Troska, J.~Mitchell,
  \href{http://www.ee.ucl.ac.uk/lcs/previous/LCS2009/LCS/lcs09_26.pdf}{Passive
  Optical Networks for Particle Physics Applications}, London Communications
  Symposium.

\bibitem{papakonstantinou2011fully}
I.~Papakonstantinou, C.~Soos, S.~Papadopoulos, S.~D{\'e}traz, C.~Sigaud,
  P.~Stejskal, S.~Storey, J.~Troska, F.~Vasey, I.~Darwazeh,
  \href{http://cds.cern.ch/record/1399721/files/05876285.pdf}{A fully
  bidirectional optical network with latency monitoring capability for the
  distribution of timing-trigger and control signals in high-energy physics
  experiments}, IEEE Transactions on Nuclear Science 58~(4) (2011) 1628--1640.

\bibitem{mitra2016common}
J.~Mitra, S.~Khan, S.~Mukherjee, R.~Paul, {for the ALICE collaboration},
  \href{http://iopscience.iop.org/article/10.1088/1748-0221/11/03/C03021}{Common
  Readout Unit (CRU)-A new readout architecture for the ALICE experiment},
  Journal of Instrumentation 11~(03) (2016) C03021.

\bibitem{mendes2017pon}
E.~Mendes, S.~Baron, D.~Kolotouros, C.~Soos, F.~Vasey,
  \href{http://stacks.iop.org/1748-0221/12/i=02/a=C02041}{The 10G TTC-PON:
  challenges, solutions and performance}, Journal of Instrumentation 12~(02)
  (2017) C02041.

\bibitem{native_phy_arria}
{Altera Corporation},
  \href{https://www.altera.com/en_US/pdfs/literature/hb/arria-10/ug_arria10_xcvr_phy.pdf}{Arria
  10 Transceiver PHY User Guide}, Arria 10 Device Handbook (2015) 615.

\bibitem{ste09meta}
J.~Stephenson, D.~Chen, R.~Fung, J.~Chromczak,
  \href{https://www.altera.com/en_US/pdfs/literature/wp/wp-01082-quartus-ii-metastability.pdf}{Understanding
  metastability in FPGAs}, Altera Corporation white paper.

\bibitem{mitra2018phase}
J.~Mitra, T.~K. Nayak, \href{http://ieeexplore.ieee.org/document/8082787/}{An
  FPGA-Based Phase Measurement System}, IEEE Transactions on Very Large Scale
  Integration (VLSI) Systems 26~(1) (2018) 133--142.
\newblock \href {http://dx.doi.org/10.1109/TVLSI.2017.2758807}
  {\path{doi:10.1109/TVLSI.2017.2758807}}.

\bibitem{Adi2009}
{Walt Kester},
  \href{http://www.analog.com/media/en/training-seminars/tutorials/MT-008.pdf}{Converting
  Oscillator Phase Noise to Time Jitter}, Analog Devices MT-008 Tutorial (2009)
  1--10.

\bibitem{Davis2007a}
P.~J. Davis, P.~Rabinowitz,
  \href{http://store.doverpublications.com/0486453391.html}{Methods of
  numerical integration}, Dover Publications, 2007.

\bibitem{neil2003understanding}
R.~Neil,
  \href{http://literature.cdn.keysight.com/litweb/pdf/5988-6254EN.pdf}{Understanding
  Jitter and Wander Measurement and Standards}, Agilent Technologies, Feb.

\bibitem{arria16intel}
{Intel Corporation},
  \href{https://www.altera.com/en_US/pdfs/literature/hb/arria-10/a10_datasheet.pdf#page=28}{Arria
  10 Device}, {A10-Datasheet} (2016) 28.

\bibitem{si5344pll}
{SILICON LABS},
  \href{https://www.silabs.com/documents/public/data-sheets/Si5345-44-42-D-DataSheet.pdf}{Si5345/44/42
  Rev D Data Sheet}.

\bibitem{bellato2014pcie}
M.~Bellato, G.~Collazuol, I.~D'Antone, P.~Durante, D.~Galli, B.~Jost, I.~Lax,
  G.~Liu, U.~Marconi, N.~Neufeld, et~al.,
  \href{http://iopscience.iop.org/article/10.1088/1742-6596/513/1/012023/meta}{A
  PCIe Gen3 based readout for the LHCb upgrade}, in: Journal of Physics:
  Conference Series, Vol. 513, IOP Publishing, 2014, p. 012023.

\bibitem{Schnecker2009Jitter}
M.~Schnecker,
  \href{http://cdn.teledynelecroy.com/files/whitepapers/designcon2009_lecroy_jitter_transfer_measurement_in_clock_circuits.pdf}{Jitter
  Transfer Measurement in Clock Circuits}, LeCroy Corporation, DesignCon.

\bibitem{jubin_link_2017}
J.~Mitra, S.~A. Khan, \href{https://gitlab.cern.ch/jmitra/GBT_BER_ARRIA10}{GBT
  BER measurement \& Transceiver Optimization - For Arria10 FPGA Development
  board}, {CRU Internal Technical Note}.

\bibitem{rudd2000statistical}
J.~Rudd,
  \href{http://notes-application.abcelectronique.com/003/3-5321.pdf}{Statistical
  confidence levels for estimating error probability}, Maxim Engineering
  Journal 37 (2000) 12--15.

\bibitem{detraz2009fpga}
S.~Detraz, C.~Sigaud, S.~Seif El~Nasr, I.~Papakonstantinou, S.~Papadopoulos,
  H.~Versmissen, P.~Moreira, C.~Soos, P.~Stejskal, S.~Silva, et~al.,
  \href{http://cds.cern.ch/record/1236362/files/p636.pdf?version=1}{FPGA-based
  bit-error-rate tester for SEU-hardened optical links}, JINST.

\bibitem{soos2008gbt}
{Csaba SOOS},
  \href{https://indico.cern.ch/event/44442/contributions/1100660/attachments/942492/1336762/Implementation_of_GBT_on_Xilinx_LHCb_meeting.pdf#page=14}{GBT
  protocol implementation on GBT protocol implementation on Xilinx FPGAs},
  {LHCB meeting}.

\bibitem{schwemmer2014evaluation}
R.~Schwemmer, J.~Cachemiche, N.~Neufeld, C.~Soos, J.~Troska, K.~Wyllie,
  \href{http://iopscience.iop.org/article/10.1088/1748-0221/9/03/C03030/pdf}{Evaluation
  of 400 m, 4.8 Gbit/s Versatile Link lengths over OM3 and OM4 fibres for the
  LHCb upgrade}, Journal of Instrumentation 9~(03) (2014) C03030.

\bibitem{altera2012xcvr}
{\relax Altera}.~Corporation,
  \href{https://www.altera.com/content/dam/altera-www/global/en_US/pdfs/literature/wp/wp-01130-stxv-transceiver.pdf}{Extending
  Transceiver Leadership at 28 nm}, White Paper.

\bibitem{altera2015hps}
{Altera Corporation},
  \href{https://www.altera.com/content/dam/altera-www/global/en_US/pdfs/literature/an/an678.pdf}{High-Speed
  link tuning using signal conditioning circuitry in Stratix V transceivers },
  White Paper.

\bibitem{alteraBlake}
{Bob Blake, Altera Corporation},
  \href{https://www.design-reuse.com/articles/11955/ensuring-serial-protocol-signal-integrity-with-fpgas-and-embedded-transceivers.html}{Ensuring
  Serial Protocol Signal Integrity with FPGAs and Embedded Transceivers}, {Web
  Article: Design \& Reuse}.

\bibitem{emphasis2010altera}
{\relax Altera}.~Corporation,
  \href{https://www.altera.com/content/dam/altera-www/global/en_US/pdfs/literature/an/an602.pdf}{Understanding
  the Pre-Emphasis and Linear Equalization Features in Stratix IV GX Devices},
  Application Note.

\bibitem{alice_cru_twiki}
{CRU Team Members},
  \href{https://twiki.cern.ch/twiki/bin/view/ALICE/CruHwFwSwDev}{ ALICE CRU
  Hardware, Firmware, and Software Development--ALICE Web TWiki} (2018).

\bibitem{Stott2011}
E.~Stott, P.~Y.~K. Cheung,
  \href{http://ieeexplore.ieee.org/document/6044838/}{Improving FPGA
  reliability with wear-levelling}, Proceedings - 21st International Conference
  on Field Programmable Logic and Applications, FPL 2011 (2011) 323--328\href
  {http://dx.doi.org/10.1109/FPL.2011.65} {\path{doi:10.1109/FPL.2011.65}}.

\bibitem{wang2002capturing}
M.~Wang, A.~Ailamaki, C.~Faloutsos,
  \href{https://www.sciencedirect.com/science/article/pii/S0166531602001086}{Capturing
  the spatio-temporal behavior of real traffic data}, Performance Evaluation
  49~(1-4) (2002) 147--163.

\end{thebibliography}
\end{document}